\documentclass[acmtog]{acmart}
\acmSubmissionID{1264}

\AtBeginDocument{%
  \providecommand\BibTeX{{%
    \normalfont B\kern-0.5em{\scshape i\kern-0.25em b}\kern-0.8em\TeX}}}

\copyrightyear{2025}
\acmYear{2025}
\setcopyright{rightsretained}
\acmConference[SA Conference Papers '25]{SIGGRAPH Asia 2025 Conference Papers}{December 15--18, 2025}{Hong Kong, Hong Kong}
\acmBooktitle{SIGGRAPH Asia 2025 Conference Papers (SA Conference Papers '25), December 15--18, 2025, Hong Kong, Hong Kong}\acmDOI{10.1145/3757377.3763854}
\acmISBN{979-8-4007-2137-3/2025/12}

\setcitestyle{square}

\usepackage{multirow}

\usepackage{pifont}
\newcommand{\xmark}{\ding{55}}%

\usepackage{multirow} 
\usepackage{url} 
\usepackage{subcaption}
\usepackage{colortbl}
\usepackage{epigraph}
\usepackage{enumitem}
\usepackage{algorithm}
\usepackage{makecell} 

\newcommand{\camr}[1]{{\color[rgb]{0,0,0}#1}}
\begin{document}

\title{Audio Driven Real-Time Facial Animation for Social Telepresence}


\author{Jiye Lee}
\email{kay2353@snu.ac.kr}
\authornote{Work done during internship at Codec Avatars Lab, Meta.}
\affiliation{%
  \institution{Seoul National University}
  \country{South Korea}
}

\author{Chenghui Li}
\email{leochli@meta.com}
\affiliation{%
  \institution{Codec Avatars Lab, Meta}
  \country{USA}
}

\author{Linh Tran}
\email{linhtran@meta.com}
\affiliation{%
  \institution{Codec Avatars Lab, Meta}
  \country{USA}
}

\author{Shih-En Wei}
\email{swei@fb.com}
\affiliation{%
  \institution{Codec Avatars Lab, Meta}
  \country{USA}
}

\author{Jason Saragih}
\email{jsaragih@meta.com}
\affiliation{%
  \institution{Codec Avatars Lab, Meta}
  \country{USA}
}

\author{Alexander Richard}
\email{richardalex@meta.com}
\affiliation{%
  \institution{Codec Avatars Lab, Meta}
  \country{USA}
}

\author{Hanbyul Joo}
\email{hbjoo@snu.ac.kr}
\authornote{Co-corresponding authors.}
\affiliation{%
  \institution{Seoul National University}
  \country{South Korea}
}

\author{Shaojie Bai}
\email{shaojieb@gmail.com}
\authornotemark[2]
\affiliation{%
  \institution{Codec Avatars Lab, Meta}
  \country{USA}
}

\renewcommand{\shortauthors}{Lee et al.}


\begin{abstract}
We present an audio-driven real-time system for animating photorealistic 3D facial avatars with minimal latency, designed for social interactions in virtual reality for anyone.
Central to our approach is an encoder model that transforms audio signals into latent facial expression sequences in real time, which are then decoded as photorealistic 3D facial avatars.
Leveraging the generative capabilities of diffusion models, we capture the rich spectrum of facial expressions necessary for natural communication while achieving real-time performance (<15ms GPU time).
Our novel architecture minimizes latency through two key innovations: an online transformer that eliminates dependency on future inputs and a distillation pipeline that accelerates iterative denoising into a single step.
We further address critical design challenges in live scenarios for processing continuous audio signals frame-by-frame while maintaining consistent animation quality.
The versatility of our framework extends to multimodal applications, including semantic modalities such as emotion conditions and multimodal sensors with head-mounted eye cameras on VR headsets.
Experimental results demonstrate significant improvements in facial animation accuracy over existing offline state-of-the-art baselines, achieving 100 to 1000$\times$ faster inference speed.
We validate our approach through live VR demonstrations and across various scenarios such as multilingual speeches. 
\end{abstract}


%


\begin{teaserfigure}
  \centering
  \includegraphics[width=1.0\textwidth, trim={0 0.5cm 0 0}]{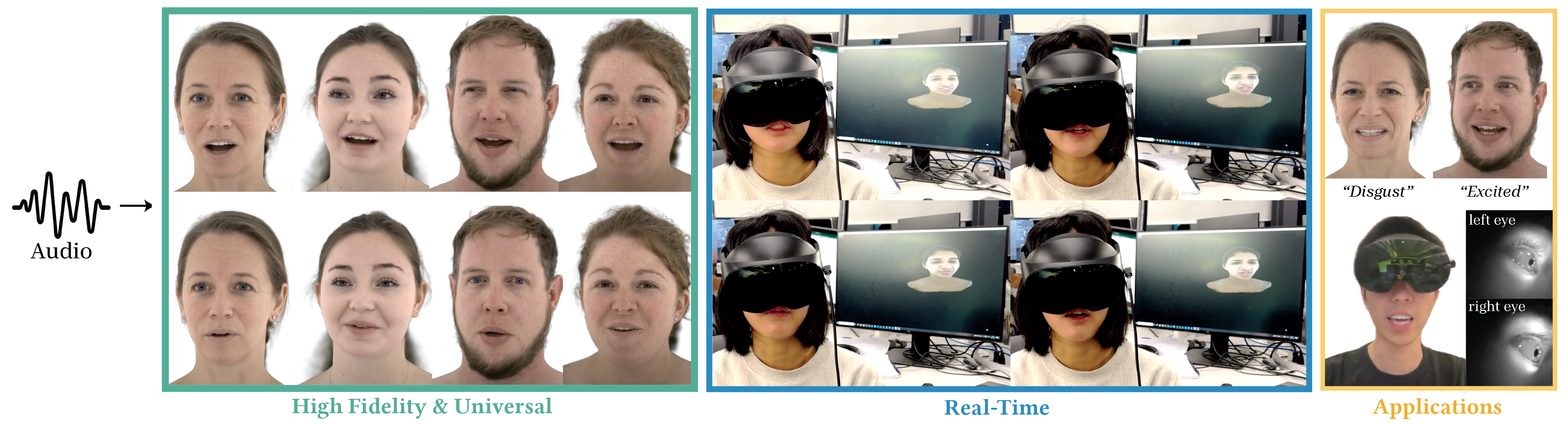}
  \caption{
 We present a system that drives \textcolor[RGB]{85,173,151}{\textbf{high fidelity and universal}} 3D facial avatars \textcolor[RGB]{49,137,190}{\textbf{real-time}} from audio.
 Our system can be extended to \textcolor[RGB]{237,174,44}{\textbf{multimodal applications}} as well.
 Project Page:
\url{https://jiyewise.github.io/projects/AudioRTA}
 }
  \label{fig:teaser}
\end{teaserfigure}

\maketitle

\section{Introduction}

The goal of photorealistic telepresence is to enable social communication in virtual spaces as naturally and seamlessly as in real life. For effective telepresence, three critical elements must converge: each participants' motion must be accurately captured and reconstructed in a \emph{photorealistic} manner to convey subtle micro-expressions; expressions must be transmitted in \emph{real-time} with minimal latency, as participants' actions must adapt instantly to the responses or expressions of their conversational partners; and the system must be highly \emph{universal}, applicable to all users regardless of individual characteristics.

Real-time performance is particularly crucial, given the dynamic and interactive nature of social communications where participants continuously respond to each other.
Achieving this real-time goal while maintaining photorealism is challenging due to sensory constraints of wearable VR devices. First, camera-based methods suffer from severe occlusions and limited viewpoints. Second, such approaches are hardware-intensive and increasingly impractical for wearable devices due to thermal and computational constraints. As VR devices become slimmer and more lightweight, e.g., towards smart glasses, these camera-based limitations become even more pronounced. 
In contrast, the audio modality is free from such sensory constraints, and as a primary medium for communication, audio contains sufficient information required for facial expression generation.

Despite the advantages of audio, existing audio-driven facial animation methods fall short of achieving photorealistic telepresence. 
Previous approaches~\cite{richard2021meshtalk, xing2023codetalker, fan2022faceformer} have explored audio-driven deformations of 3D facial template meshes. 
These representations often lack sufficient detail for conveying subtle facial cues essential in social interactions. 
Additionally, directly modeling mesh deformations may lead to computational challenges that limit real-time performance.
More recent approaches synthesize high-fidelity avatars from audio~\cite{ng2024audio2photoreal, li2025talkinggaussian}, yet primarily operate in an offline manner requiring entire audio sequences as input rather than processing audio streams in real-time, also with minimal consideration for computational latency.

We present a real-time system for audio-driven facial telepresence in VR that achieves high-fidelity, universal, and real-time performance simultaneously. \camr{Notably, our system aims to address the computational and latency challenges of audio-driven facial animation generation.}
Our system's core is an encoder model generating latent facial expressions from audio in real-time, which are subsequently decoded into universal and photorealistic 3D avatars~\cite{li2024uravatar}. 
While built upon diffusion models to capture rich facial expressions, we introduce novel architectural elements to meet real-time constraints: an online transformer design and training to eliminate dependency on future audio inputs and a distillation pipeline accelerating iterative denoising into a single step, achieving real-time performance (100FPS in GPU time). We further address system design challenges for live-driving scenarios and demonstrate multimodal extensions for diverse telepresence applications.

Our contributions can be summarized as follows:
(1) a novel system architecture for real-time facial expression generation via diffusion models with online Transformer and distillation-based acceleration;
(2) live-driving system design and demonstrations;
(3) multimodal applications including emotional and multi-sensor inputs using a VR headset.

\section{Related Work}
\label{sec:relatedwork}

\subsection{Audio Driven Face Generation}
The goal of audio driven face generation is to create synchronized facial movements, either as video or 3D sequences, by taking an audio sequence as input. 
Synthesizing portrait image sequences of a person talking in sync with audio~\cite{suwajanakorn2017synthesizingobama, jamaludin2019yousaidthat, prajwal2020wav2lip, zhou2020makeittalk} has been an active area of research in computer vision and graphics. 
Since methods that rely solely on 2D images lack 3D structural information, leveraging 3D representations~\cite{zhang2023sadtalker, ye2024real3dportrait, ji2021audioemo} has been explored for improved spatiotemporal consistency. 
More recently, advancements in diffusion generative models have enabled diffusion-based approaches~\cite{xu2024vasa, shen2023difftalk, stypulkowski2024diffusedgan, xu2024hallo, cui2024hallo2} to synthesize high-quality portrait images synchronized with audio.

To overcome limitations of 2D generation such as lack of structural information or limitation to a single viewpoint, recent approaches focus on synthesizing 3D faces from audio. 
One line of research focuses on generating 3D facial avatars by directly deforming geometry.
A widely explored direction is to generate mesh deformations from audio, given a predefined 3D face mesh template~\cite{richard2021meshtalk, cudeiro2019voca, nocentini2025scantalk, fan2022faceformer, xing2023codetalker, haque2023facexhubert, karras2017audiodriven, thambiraja2023imitator}. These methods, however,
lack the capability for high-fidelity expression rendering.
Recent advancements in neural 3D representations, particularly Neural Radiance Fields (NeRF), have been explored to render high-fidelity audio-driven faces~\cite{li2023efnerf, liu2022sspnerf, yao2022dfanerf, ye2023geneface, guo2021ad, shen2022dfrf, ye2023geneface++}. 
More recently, methods utilizing Gaussian Splatting (3DGS)~\cite{kerbl20233dgs} have been proposed to model detailed geometry and appearance deformations~\cite{li2025talkinggaussian, chen2024gstalker, aneja2024gaussianspeech, he2024emotalk3d}. While these approaches achieve high-fidelity face generation, they require the network to learn deformations directly from audio, which is hard to be generalized across identities with diverse facial geometries and 
leading to computational overhead which makes such approaches unsuitable for real-time applications. 
\camr{GSTalker~\cite{chen2024gstalker} and GaussianTalker~\cite{cho2024gaussiantalker, yu2024gstalker}
address computation issue by real-time rendering through an optimized 3D representation based on 3DGS, but are trained on personalized data and thus generalization to multiple identities is limited.
Moreover, while these methods achieve real-time rendering, the audio-based deformations are computed in an offline manner, taking the entire audio sequence as input.}

An alternative direction, where our work falls into, employs an encoder-decoder structure where audio input generates latent facial expressions which are decoded into a 3D representation.
In the 2D domain, Vasa-1~\cite{xu2024vasa} enables online, real-time performance by producing low dimension latent vectors from audio and utilizing an image decoder to synthesize the final output.
Regarding 3D, 
recent methods like TalkShow~\cite{yi2023talkshow} use convolution-based regression models to predict expression parameters that are decoded with FLAME~\cite{FLAME:SiggraphAsia2017} into facial mesh sequences.
Due to recent advances in diffusion models, 
FaceTalk~\cite{aneja2024facetalk} and DiffPoseTalk~\cite{sun2024diffposetalk}
employs a universal diffusion-based model for expression generation. FaceTalk uses Neural Parametric Head Models (NPHM)~\cite{giebenhain2023nphm} for decoding expressions into detailed meshes, while DiffPoseTalk further leverages GaussianAvatars~\cite{qian2024gaussianavatars} for high-fidelity rendering.
Audio2Photoreal~\cite{ng2024audio2photoreal} also uses diffusion models and a photorealistic avatar decoder but is limited to personalized conversational settings with dual audio inputs.
Such methods can not operate in real-time due to their offline architecture and inherently slow denoising process of diffusion models.

\camr{Another relevant research direction is to incorporate emotional expressiveness into audio-driven 3D facial animation~\cite{danvevcek2023emotional, peng2023emotalk, wu2024probtalk3d, zhao2024media2face} using emotion labels as conditional input. However, these template mesh-based methods struggle to capture subtle emotional nuances. While EmoTalk3D~\cite{he2024emotalk3d} achieves higher fidelity through novel-view synthesis, existing approaches remain computationally expensive and lack real-time performance or cross-identity generalization.}

\subsection{Telepresence with Photorealistic Facial Avatars}

The generation of photorealistic digital avatars has been a long-researched topic in computer graphics. Advances in multi-view capture systems have facilitated high-fidelity reconstruction appearance and geometry of individuals~\cite{beeler2011highquality, fyffe2014drivinghigh, kirschstein2023nersemble, joo2018totalcapture}. 
Recently, methods on avatar generation have shifted their focus on leveraging deep learning, 
using volumetric representations~\cite{cao2022authentic, rosu2022neuralstrands}, radiance fields~\cite{kirschstein2023nersemble}, and Gaussian Splatting~\cite{saito2023rgca, li2024uravatar}. Such learning-based approaches also allow avatar generation in much more constrained setups, such as capture from a hand-held phone camera~\cite{cao2022authentic, athar2024bridgingthegap}, casual few-shot images~\cite{buehler2024cafca} 
and with on-device computation~\cite{ma2021pixel, saito2024squeezeme}.

For telepresence, the ability to drive the avatar in a human-like manner is as important as its generation, whereas research has focused on utilizing wearable devices which is challenging due to the limited observations of the driving user's movement.
Several methods address this by using additional sensors on wearable devices (e.g., VR headset) with a protruding mount~\cite{li2015facial, htc_vive} or in a third-person view~\cite{thies2018facevr}. 
~\citet{wei2019VR} and ~\citet{schwartz2020eyes} improved hardware stability by attaching mouth cameras right below a consumer-level VR headset, but are still limited to personalized use. ~\citet{bai2024ue} used only 4 built-in head-mounted cameras (HMCs) on a consumer-level VR headset to robustly track facial expressions real-time in an universal setting. 
Very recently, ~\citet{tran2024voodooxp} also showed real-time reenactment of a single portrait image using a consumer-level VR headset.

\section{System Overview}
\label{sec:codec}
%


We propose a framework to generate facial expressions in real time from raw audio signals. 
Rather than directly modeling facial geometry deformations from audio, our method represents facial expressions as latent expression codes. 
The encoder $\mathcal{E}$ generates latent expressions from audio signals in real-time, which are subsequently transformed into photorealistic facial avatars via the decoder $\mathcal{D}$.

The encoder $\mathcal{E}$ serves as the primary component of our system, functioning as a sequence-to-sequence model which takes a sequence of audio signals $\mathbf{W} = \{ \mathbf{w}_t \}_{t=0}^T$ as input and outputs a sequence of latent facial expressions $\mathbf{X} = \{ \mathbf{x}_t \}_{t=0}^T$ and gaze $\mathbf{G} = \{ \mathbf{g}_t \}_{t=0}^T$. At time $t$, $\mathbf{x}_t \in \mathbb{R}^{256}$ indicates the latent expression code and $\mathbf{g}_t \in \mathbb{R}^{2 \times 3}$ is the direction of both eyes in unit vectors.
Notably, to ensure real-time performance, the encoder model $\mathcal{E}$ is designed both to reduce computational latency and to eliminate any lookaheads. This indicates that although the encoder produces a full sequence of outputs for the entire duration, each output element corresponding to time $t$ is computed solely from the information available in the input sequence from time range 0 through $t$, with no future data from $t+1$ to $T$ incorporated into the computation. 

The latent expressions $\mathbf{X}$ which the encoder $\mathcal{E}$ outputs are defined in the latent space built upon the Universal Relightable Prior Model~\cite{li2024uravatar} framework.
In the framework, a variational autoencoder (VAE) based network is employed to learn a shared latent distribution of expressions across identities from multi-view facial images with large-scale multi-identity data. 
Consequently, the expressions defined in the shared latent space contains semantic nuances common to various identities; a single expression can drive multiple identities, ensuring universal applicability.

\begin{figure}
\includegraphics[width=0.9\linewidth, trim={0 0.5cm 0 0}]{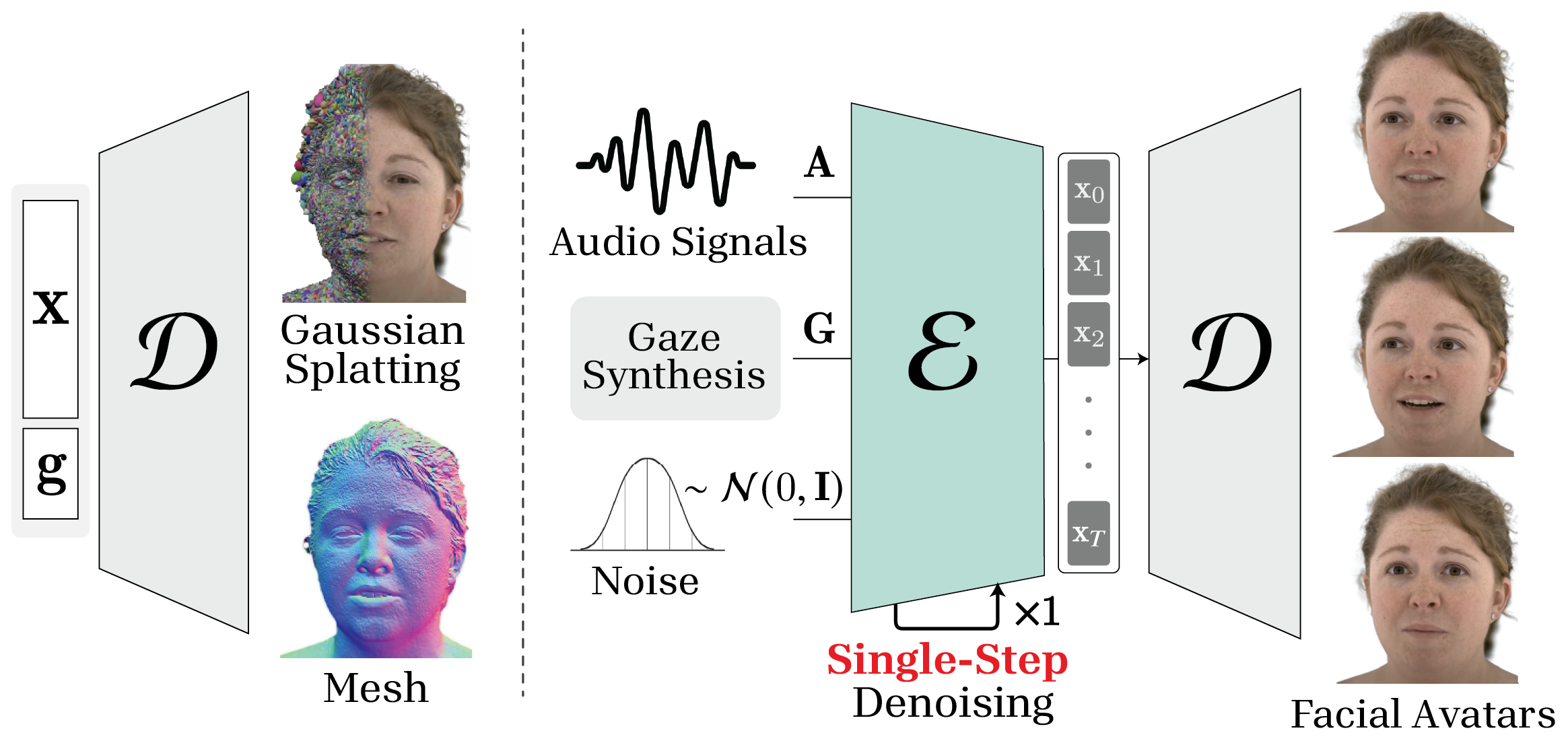}
  \caption{
  Overview of the decoder (left) and the encoder-decoder pipeline (right). The encoder $\mathcal{E}$ generates expression codes in real time based on single-step denoising of a diffusion model, which is decoded into 3DGS and mesh by the decoder $\mathcal{D}$.} 
  \label{fig:sys_overview}
\end{figure}

The decoder $\mathcal{D}$ in our system is adapted from the decoder in the Universal Relightable Prior Model~\cite{li2024uravatar} framework, where the decoder generates photorealistic avatars from the given facial expressions.
The decoder takes as input the face expression code $\mathbf{x}$, gaze $\mathbf{g}$, and id-specific bias maps $\Theta_{\text{id}}$ and produces face mesh $\mathbf{M}$ and a set of 3D Gaussians
$\mathcal{S}  = \{ \mathbf{s}_k \}$.
Each Gaussian $\mathbf{s}_k$ includes translation, rotation, scale, opacity, and color attributes.
To decode the identity-agnostic expressions into person-specific details of the facial avatars, the decoder 
is parameterized by $\Theta_{\text{id}}$ which are computed from a mean texture map and a mean geometry map.
This decoding process can be written as $\mathbf{M}, \mathcal{S} \leftarrow \mathcal{D}(\mathbf{x}, \mathbf{g}; \Theta_{\text{id}})$.
%
%
We refer interested readers to the Universal Relightable Prior Model~\cite{li2024uravatar} for details of computing the latent space and the decoder $\mathcal{D}$.
Fig.~\ref{fig:sys_overview} describes the overall pipeline.

\section{Audio Driven Real-Time Encoder}
\label{sec:audioua}


\begin{figure}
  \includegraphics[width=1.0\linewidth, trim={0 0.5cm 0 0}]{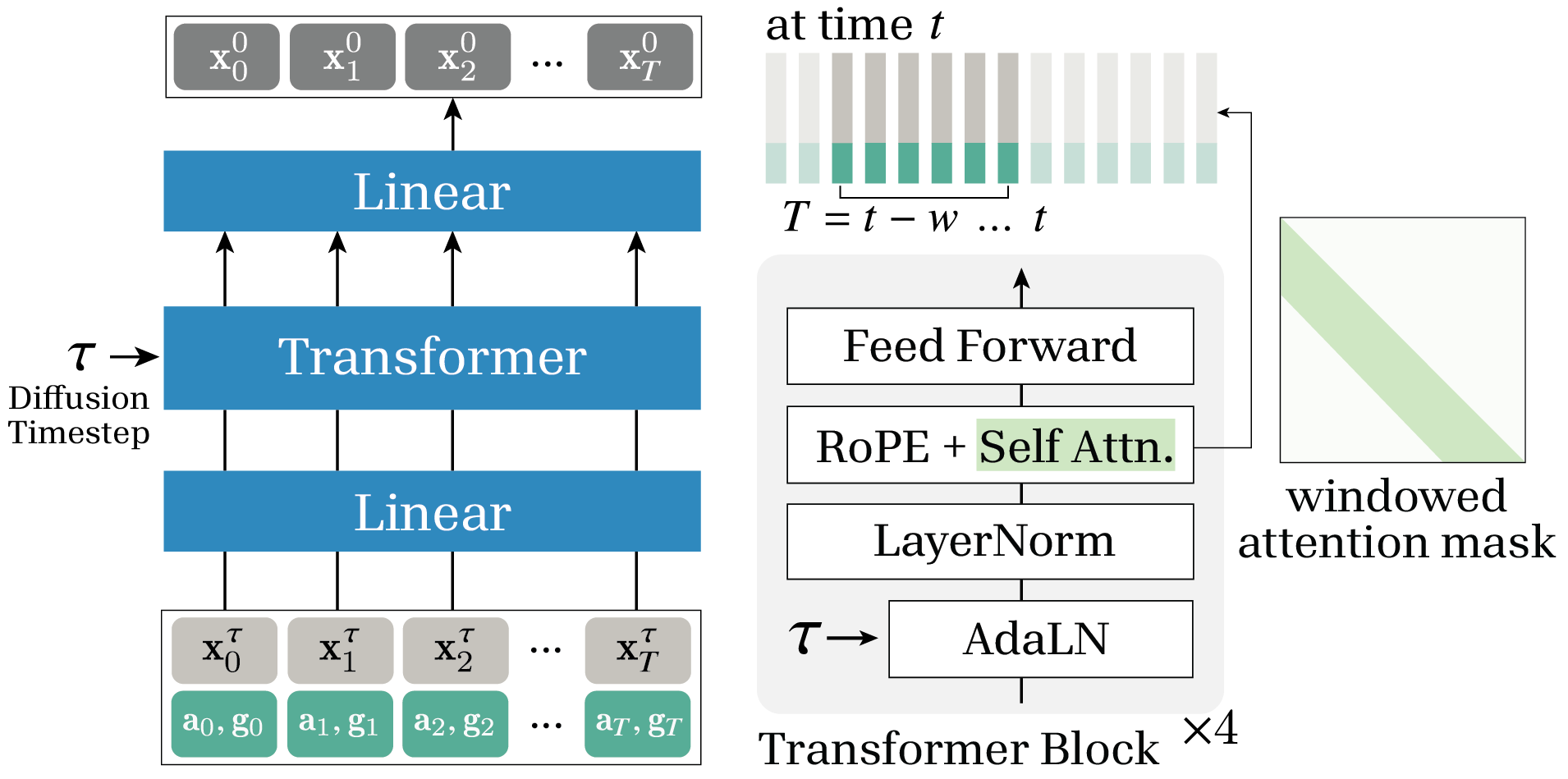}
  \caption{Transformer-based denoising network architecture. Windowed attention mask is applied in the self-attention layer of each Transformer block.}
  \label{fig:diffusion_arch}
\end{figure}

\subsection{Diffusion-based Expression Generation}
\label{sec:audio_diffusion}
The encoder $\mathcal{E}$ consists of a neural network that generates latent expressions from audio signals. We adopt diffusion models for their ability to generate natural and expressive facial animations.
We follow the diffusion definition presented by ~\citet{ho2020ddpm}, which involves two processes. The forward process, where noise is incrementally added to clean data $\mathbf{X}^0$ is formulated as: 
\begin{equation}
    q(\mathbf{X}^\tau|\mathbf{X}^{\tau-1}) \sim \mathcal{N}(\sqrt{\alpha_\tau}\mathbf{X}^{\tau-1}, (1-\alpha_\tau)\mathbf{I})
\end{equation}
where $\mathbf{X}^0$ indicates the sequence of facial expressions without noise, $\tau \in [1, ... , N]$ denotes the diffusion step and $\alpha_{\tau} \in (0,1)$ is defined by a fixed variance schedule. 

The reverse process reconstructs the clean sample $\mathbf{X}^0$ by denoising noise  $\mathbf{X}^N \sim \mathcal{N}(0, \mathbf{I})$. 
As the reverse process relies on the distribution $q(\mathbf{X}^{\tau-1} | \mathbf{X}^\tau)$ which is intractable, a neural network is trained to approximate this distribution.
Note that $\tau$ denotes the diffusion step (not the temporal timestep $t$), and in our setting data $\mathbf{X}$ refers to the whole facial expression sequence $\mathbf{X} = \{\mathbf{x}\}_{t=0}^T$. 

Following~\cite{tevet2023mdm, nichol2021improved}, in practice we train a denoising neural network $\mathcal{F}$ to directly predict the clean data\footnote{The hat symbol indicates that the value is an estimated output from a neural network.} $\hat{\mathbf{X}}^0$ from noisy data $\mathbf{X}^\tau$:
\begin{equation}
    \hat{\mathbf{X}}^0 = \mathcal{F}(\mathbf{X}^\tau| \tau, \mathbf{A}, \mathbf{G}) 
\end{equation}
where the conditional $\mathbf{A}=\{\mathbf{a}_t\}_{t=0}^T$ represents a sequence of audio features $\mathbf{a}_t \in \mathbb{R}^{d_a}$, extracted from the raw audio signal sequence $\mathbf{W}$ with the Wav2Vec~\cite{schneider2019wav2vec1} audio encoder, with $d_a$ indicating the feature dimension. Using a pretrained audio encoder extracts speaker-agnostic features from the raw audio signals, enabling the encoder model to adapt to unseen audio.
The conditional $\mathbf{G} = \{\mathbf{g}_t\}_{t=0}^{T}$ denotes the sequence of gaze directions. 
\camr{From the observation that gaze is correlated with speech but with relatively large randomness, i.e., a person can blink anytime when speaking,}
and that gaze is highly correlated with the movement of upper face regions, we augment the denoising network to incorporate gaze as an additional conditional input. 
During training, classifier-free guidance (CFG)~\cite{ho2022cfg} is applied by randomly dropping out conditioning $\mathbf{A}$ and $\mathbf{G}$ with a certain probability.
Since gaze direction information $\mathbf{G}$ is not available in purely audio-driven scenarios, we first synthesize $\mathbf{G}$ prior to its usage. The denoised latent facial expressions $\mathbf{X}$ and gaze $\mathbf{G}$ are concatenated before being decoded into 3D facial avatars.

\subsubsection{Architecture} 
We use the Transformer model as a denoising network for diffusion for its effectiveness in capturing intricate relationships in sequential data. 
Conditionals $\mathbf{A}$ and $\mathbf{G}$ are added by concatenating with noisy sample $\mathbf{X}^\tau$. 
The diffusion step $\tau$ information is added with an AdaLN layer in the transformer block. 

For real-time demonstrations, the model should be online; it should not refer to future information in the input sequence. 
To this end, we propose applying a windowed mask on the self-attention layers so that at time frame $t$, the attention mechanism is only applied to the sequence between $t-w$ to $t$, where $w$ is the window size.
Moreover, to make the model adaptable to arbitrary sequence lengths we apply Rotary Positional Embeddings (RoPE)~\cite{su2024roformer}, which incorporates relative position information through rotation matrix product.
Fig.~\ref{fig:diffusion_arch} shows the online Transformer architecture. 

\begin{figure}
\includegraphics[width=0.98\linewidth, trim={0 0.3cm 0 0}]{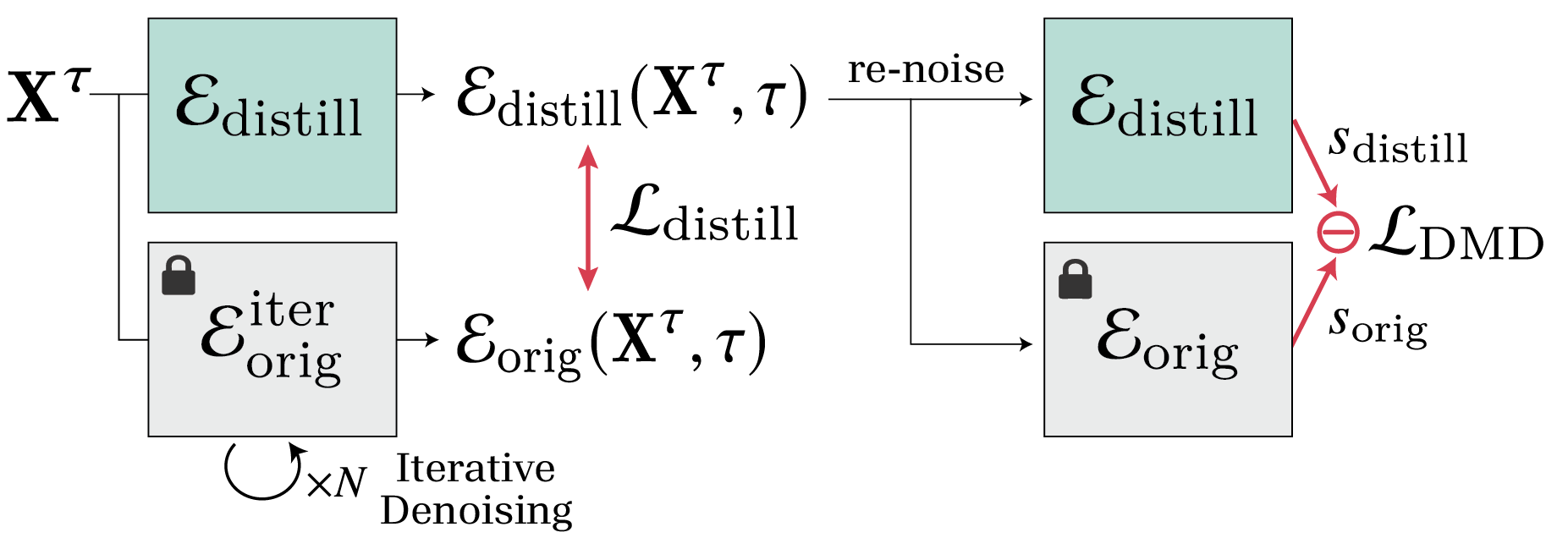}
  \caption{Pipeline of distillation training. The distilled model first learns via $\mathcal{L}_\text{distill}$ from the original model's multi step and its own single step outputs, with re-noised samples are used to compute $\mathcal{L}_\text{DMD}$.}
  \label{fig:distill_train}
\end{figure}

\subsubsection{Training Objectives} 
The denoising neural network $\mathcal{F}$ is trained to minimize loss $\mathcal{L}$, which is defined as
\begin{equation}
    \mathcal{L} = \lambda_{\text{simple}}\mathcal{L}_{\text{simple}} + \lambda_{\text{geom}}\mathcal{L}_{\text{geom}}. \label{eq:orig_diff_loss}
\end{equation}
First, we use a simplified ELBO objective~\cite{ho2020ddpm},
\begin{equation}
    \mathcal{L}_{\text{simple}} = \| \hat{\mathbf{X}}^0 - \mathbf{X}^0 \|^2
\end{equation}
where $\hat{\mathbf{X}}^0 = \mathcal{F}(\mathbf{X}^\tau | \tau, \mathbf{A}, \mathbf{G})$. This loss directly measures the $L_2$ distance of the denoised latent expression sequence predicted by the denoising network $\mathcal{F}$.

As our network operates online and relies solely on past audio without access to future information, learning temporally coherent facial expressions and lip motion is more challenging than in offline settings; thus we introduce facial geometry based training objectives to guide the network to learn more natural temporal dynamics.
Facial geometry is defined as a sequence of face meshes $\{\mathbf{M}\}_{t=0}^{T}$ obtained by decoding the expression sequence $\mathbf{X}$.\footnote{To compute the geometry losses, we include the identity information $\Theta_{\text{id}}$ in the data batch during training. While the encoder itself does not require identity information, during training the encoder-generated expressions and identity information are provided into the frozen decoder model to obtain geometry for loss computation.}
The geometry loss $\mathcal{L}_\text{geom}$ consists of a velocity loss $\mathcal{L}_{\text{vel}}$ and a jitter loss $\mathcal{L}_{\text{jitter}}$, which are measured based on the velocity and jitter of the facial mesh vertices $\mathbf{v}$.

The velocity loss $\mathcal{L}_{\text{vel}}$ measures the $L_2$ distance between the velocities of the predicted and ground truth face mesh vertices. The velocity of a vertex $\mathbf{v}_i^t$ is defined as $\dot{\mathbf{v}}_i^t = \mathbf{v}_i^t - \mathbf{v}_i^{t-1}$ for $t \geq 1$ and the loss is defined as
\begin{equation}
\mathcal{L}_{\text{vel}} = \sum_{t=1}^T \sum_{i=1}^n w_i \|\dot{\mathbf{v}}_i^t - \dot{\mathbf{v}}_i^{t,\text{gt}}\|^2
\label{eq:geom_vel_loss}
\end{equation}
where $w_i$ is the weight for the $i$-th vertex. The weight $w_i$ is set differently for mouth and non-mouth regions.

To ensure smooth facial movements, we also define jitter loss $\mathcal{L}_{\text{jitter}}$.
The jitter of a vertex is computed as the derivative of acceleration, or $\mathbf{v}_i^t - 3\mathbf{v}_i^{t-1} + 3\mathbf{v}_i^{t-2} - \mathbf{v}_i^{t-3}$ for $t \geq 3$. Instead of directly measuring the distance between predicted and ground truth jitter, we propose a normalized, ratio-based jitter loss
\begin{equation}
\mathcal{L}_{\text{jitter}} = \sum_{t=3}^T \sum_{i=1}^n \left\| \frac{\mathbf{v}_i^t - 3\mathbf{v}_i^{t-1} + 3\mathbf{v}_i^{t-2} - \mathbf{v}_i^{t-3}}{\mathbf{v}_i^{t,\text{gt}} - 3\mathbf{v}_i^{t-1,\text{gt}} + 3\mathbf{v}_i^{t-2,\text{gt}} - \mathbf{v}_i^{t-3,\text{gt}}} - 1 \right\|^2
\label{eq:geom_jitter_loss}
\end{equation}
Such normalization ensures balanced contributions across multiple identities with diverse face geometries by preventing the model from overfitting to larger faces, which would otherwise dominate the loss, while smaller faces would be underrepresented.

\subsubsection{Gaze Synthesis} 
As our system takes only audio signals as input, we also synthesize gaze movements to use as input conditioning for the diffusion model. 
Gaze specifically refers to the tracked sequence of gaze direction in unit vectors $ \mathbf{g} \in \mathbb{R}^{2\times3}$, and is not defined as eye movements such as blinking. 
For example, while blinking, gaze tracking would often fail and result in noisy or inconsistent direction vector sequences.
For gaze synthesis, we adopt a graph-based approach inspired by motion graphs. 
In this approach, nodes are constructed based on gaze vectors over a sequence of frames, including gaze position and velocity, while edges are connected based on the distance between nodes. 
This ensures plausible transitions and enables the synthesis of smooth and physically consistent gaze sequences via graph traversal. 
Further details on the approach are provided in the Supp. Sec. 3.2.
Furthermore, gaze synthesis can be easily extended to multimodal settings by replacing gaze with eye features tracked using a head-mounted camera (HMC) in a VR device.
Refer to Sec.~\ref{sec:multimodal_application} for multimodal driving applications.

\subsection{Single Step Distillation for Real-Time Generation}
\label{sec:singlestep}
To generate expressions from audio in a single step, a basic approach is to directly regress the facial expressions, but regression falls short in expressiveness.
On the other hand, diffusion models can generate realistic and expressive motions by learning the probability distribution of expressions but are inherently slow. 
To address this, we propose a single step denoising distillation to leverage the expressive prior learned by diffusion models while accelerating the slow inference process.
Given a diffusion model (Sec.~\ref{sec:audio_diffusion}) $\mathcal{E}_\text{orig}$ which reconstructs $\mathbf{X}^0$ through iterative denoising, we train a distilled model $\mathcal{E}_\text{distill}$ which learns to denoise $\mathbf{X}^0$ from $\mathbf{X}^\tau$ in a single step. 
\subsubsection{Training \& Objectives} 
 The distilled model $\mathcal{E}_\text{distill}$ share an identical network architecture with the original model $\mathcal{E}_\text{orig}$, and the parameters of the distilled model are initialized to be identical to the parameters of the original diffusion model. 
To make the distilled model learn the probabilistic priors learned by the original diffusion model, we incorporate two loss terms $\mathcal{L}_\text{distill}$ and $\mathcal{L}_\text{DMD}$ for distillation.
$\mathcal{L}_\text{distill}$ loss directly leverages the denoising capability of the original diffusion model. 
\begin{equation}
    \mathcal{L}_\text{distill} = \|\mathcal{E}_\text{distill}(\mathbf{X}^\tau, \tau) - \mathcal{E}_\text{orig}^\text{iter}(\mathbf{X}^\tau, \tau)\|^2
\end{equation}
The superscript `iter' indicates that the denoised expressions $\mathbf{X}^0$ from $\mathcal{E}_\text{orig}$ are obtained through iterative denoising. 
CFG is also applied to the original model during this process. As a result, the distilled model no longer requires CFG computation during inference, which improves computational efficiency.

$\mathcal{L}_\text{DMD}$ loss, inspired by~\citet{yin2024dmd}, enforces the generated samples of the distilled model to follow the distribution learned by the original diffusion model. The distance between distributions are measured using Kullback-Leibler (KL) divergence:
\begin{equation}
\begin{split}
        \mathcal{L}_\text{DMD} &= D_\text{KL}\big(p_\theta^{\text{distill}} \| p_\theta^{\text{orig}}\big) \\
        &= \mathbb{E} \big[ - \big(\log p_\theta^\text{distill}(\mathbf{z} ) - \log p_\theta^\text{orig}(\mathbf{z}) \big)  \big]
\end{split}
\end{equation}
where $\mathbf{z} = \mathcal{E}_\text{distill}(\mathbf{X}^\tau, \tau)$. 
Taking the gradient of $\mathcal{L}_\text{DMD}$ leads to:
\begin{equation}
    \nabla_\theta\mathcal{L}_\text{DMD} = \mathbb{E}\big[ - \big( \mathbf{s}_\text{orig}(\mathbf{z}) - \mathbf{s}_\text{distill}(\mathbf{z})\big) \nabla_\theta\mathcal{E}_\text{distlll} \big]
\end{equation}
where $\mathbf{s}_\text{orig}(\mathbf{z} ) = \nabla_{\mathbf{z}} p_\theta^\text{orig}(\mathbf{z})$ and $\mathbf{s}_\text{distill}(\mathbf{z} ) = \nabla_{\mathbf{z}} p_\theta^\text{distill}(\mathbf{z} )$ are the score functions of both distributions.
The single step prediction of the distilled model $\mathbf{z} = \mathcal{E}_\text{distill}(\mathbf{X}^\tau, \tau)$ undergoes re-noising with the noise schedule of the original model, and 
the new diffused sample is passed through $\mathcal{E}_\text{orig}$ and $\mathcal{E}_\text{distill}$ to get the score $\mathbf{s}_\text{orig}$ and $\mathbf{s}_\text{distill}$, respectively. 
Apart from the losses for learning the prior, we also add geometry loss $\mathcal{L}_\text{geom}$ as in Eq.~\ref{eq:geom_vel_loss} and Eq.~\ref{eq:geom_jitter_loss} to ensure accurate alignment to ground truth.
Refer to Fig.~\ref{fig:distill_train} for the distillation pipeline.

\subsection{System Design for Live-Driving Scenarios}
\label{sec:live_drive_system}
In this section we discuss system design to make our pipeline robust in live-driving scenarios, where audio signal is injected to the system every frame in a certain FPS and the model has to generate corresponding facial expressions.
Formally, this means that in order to produce frame $\mathbf{x}_t$ at time step $t$, the model receives a waveform sequence $\{\mathbf{w}\}_{0:t}$ up to time $t$ as well as randomly sampled Gaussian noise $\mathcal{N}(0,\mathbf{I})$ as input, but does not have access to any future information beyond frame $t$.

\subsubsection{Audio Encoder with Causality} The raw audio waveform sequence is first encoded to audio features using an pretrained audio encoder. 
In live driving scenarios, where future audio information is unavailable, the \emph{causality} of the audio encoder, i.e., its ability to operate with zero lookahead, should be ensured.
While numerous pretrained audio encoders exist, many rely on future information such as Wav2vec 2.0~\cite{baevski2020wav2vec2} with non-causal convolution layers and HuBERT~\cite{hsu2021hubert} which uses self-attention on future context. We adopt Wav2vec 1.0~\cite{schneider2019wav2vec1} audio encoder, where causality is preserved by causal convolution layers. 
Refer to Supp. Sec. 4 for the details on the audio encoders.

\subsubsection{Outpainting for Consistency} 
Due to the stochastic nature of the diffusion model's denoising process, when sampling is done at every frame $t$ the sampled expressions $\mathbf{x}_t$ may be distinct among frames.
To enforce temporal consistency, we utilize the image outpainting technique commonly used in image diffusion, which preserves the masked regions while filling in the new regions to maintain coherence.
The outpainting process can be expressed as
\begin{equation}
    \{\mathbf{x}^{\tau}\}^\text{outpaint}_{0:t} = m \odot \{\mathbf{x}^\tau\}_{0:t} + (1-m) \odot \big(\{\mathbf{x}^\tau\}_{0:t-1} \oplus \mathbf{0}\big),
\end{equation}
where $\{\mathbf{x}^\tau\}_{0:t}$ and $\{\mathbf{x}^\tau\}_{0:t-1}$ are derived from denoising network output $\{\hat{\mathbf{x}}^0\}_{0:t}$ and $\{\hat{\mathbf{x}}^0\}_{0:t-1}$
at frame $t$ and $t-1$ respectively at diffusion timestep $\tau$. The mask $m$ is set to zero for frame $0$ to $t-1$ and 1 at time $t$. $\{\mathbf{x}^\tau\}_{0:t-1} \oplus \mathbf{0}$ indicates that the zero vector $\mathbf{0}$ is concatenated to expand the sequence to length $t$. For the distillation model, the denoising process is set to a single step, the outpainting process is simplified to $\tau = N$.

\section{Multimodal Applications}
\label{sec:applications}
\begin{figure}
\includegraphics[width=1.0\linewidth, trim={0 0.4cm 0 0.0cm}]{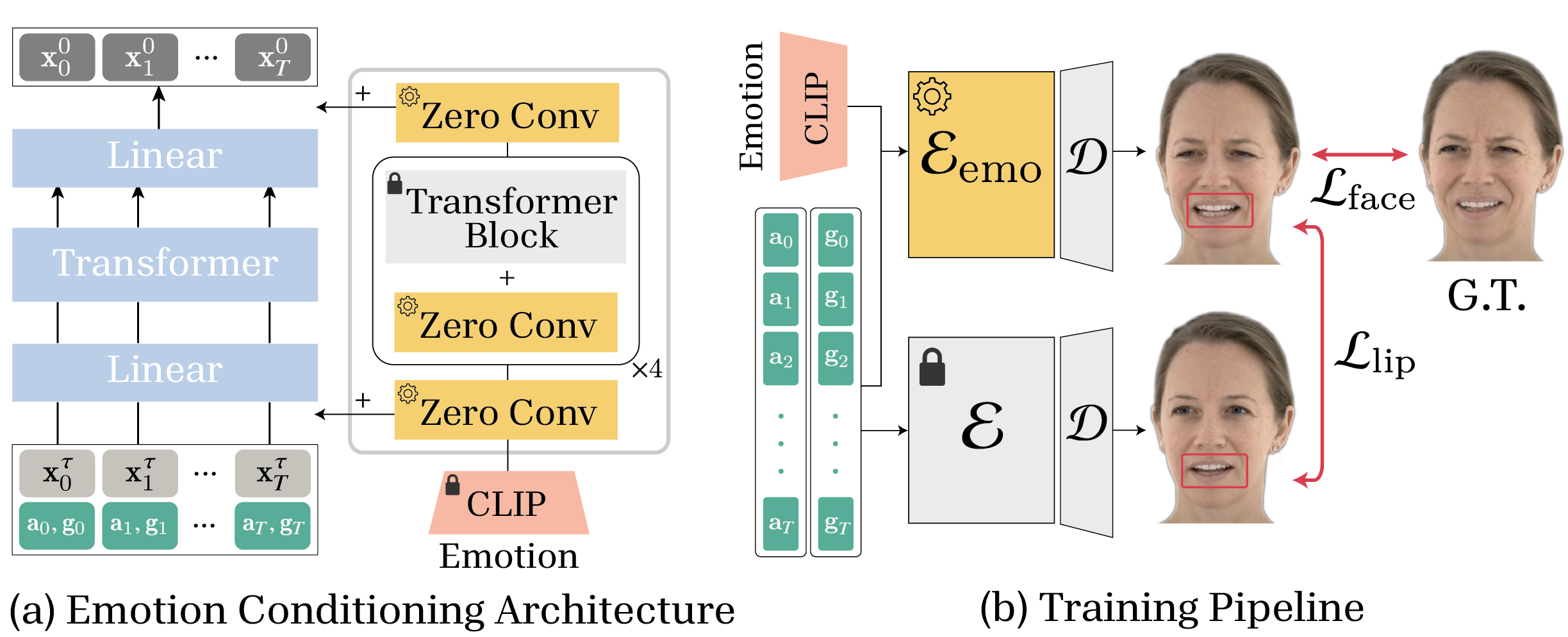}
  \caption{Emotion conditioning architecture (left) and training pipeline (right). 
  Zero convolutional layers are added between the linear and Transformer blocks for conditioning. These layers are trained with $\mathcal{L}_\text{face}$ from emotion-conditioned model and $\mathcal{L}_\text{lip}$
from the neutral model's lip motion.}
  \label{fig:emotion_cond_arch}
\end{figure}
In this section we demonstrate multimodal extensions of our audio driven pipeline by integrating additional conditional inputs with different modalities, which can be adapted to diverse social telepresence scenarios.
\subsection{Emotion Conditioning}
\label{sec:emo_cond}

Leveraging the versatility of generative models, our approach can naturally be extended to semantic modalities such as emotions.
We extend the model to learn emotional expressions by initially training the diffusion model on abundant non-emotional audio-to-expression data and subsequently finetuning with emotion-labeled data.

\subsubsection{Architecture}  
Instead of directly fine-tuning the pretrained diffusion model $\mathcal{E}$, we freeze the pretrained model and incorporate trainable zero-convolutional layers between the transformer blocks of the frozen model to learn emotional context while preserving the learned priors (e.g., lip synchronization).
Zero convolution layers are $1 \times 1$ convolutional layers where weights and biases are initialized as zeros; before the parameters are updated, the output of the finetuned model is identical to the original model.
Given an emotion label, the corresponding embedding $\mathbf{c} \in \mathbb{R}^{512}$ is extracted using a pretrained CLIP model. The CLIP embeddings are then passed through a convolution layer $\mathcal{Z}$ with zero-initialized trainable parameters $\Theta_z$. Let $\mathbf{y}$ represent the output of a specific layer in the original model. The emotion embedding $\mathbf{c}$ passes through the zero convolution layer $\mathcal{Z}$, and its output is added to $\mathbf{y}$ to produce the conditioned output $\mathbf{y}'$. This process is formally expressed as $\mathbf{y}' = \mathbf{y} + \mathcal{Z}(\mathbf{c}; \Theta_z)$.

\subsubsection{Training Objectives.}  
The emotion-conditioned model $\mathcal{E}_\text{emo}$ extends the neutral model $\mathcal{E}$ by adding zero convolution layers $\{\mathcal{Z}\}$.
During training, only the parameters $\{\Theta_z\}$ of $\{\mathcal{Z}\}$ are optimized, while $\mathcal{E}$ is frozen. 
The goal is to enable $\mathcal{E}_\text{emo}$ to produce emotion-conditioned expressions $\mathbf{X}_\text{emo}^0$ that capture emotion-specific facial expressions (e.g., frowning, raising eyebrows) while preserving accurate lip synchronization guided by $\mathcal{E}$. 
To this end,  
the layers $\{\mathcal{Z}\}$ are trained with two loss terms: $\mathcal{L}_\text{face}$, which is applied to the face meshes decoded from the output of $\mathcal{E}_\text{emo}$, and $\mathcal{L}_\text{lip}$, which is applied to the lip region of the facial meshes obtained from the neutral model $\mathcal{E}$.
Refer to Fig.~\ref{fig:emotion_cond_arch} (right) for the training pipeline.

\subsection{Multimodal Sensors}
\label{sec:multimodal_application}
\begin{figure}
\includegraphics[width=0.9\linewidth,  trim={0 0.5cm 0 0}]{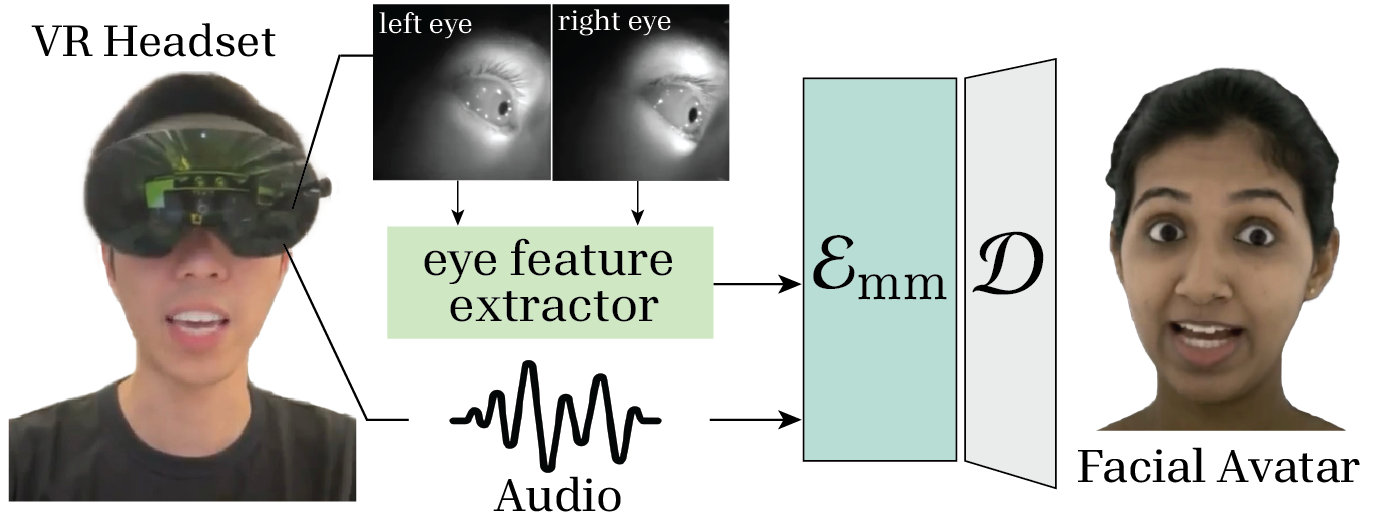}
  \caption{Pipeline for multimodal applications using a VR headset, with 2 HMC eye cameras and a microphone. Eye features are extracted from HMC images, and are given as a input with the audio signal.}
  \label{fig:mm_sys}
\end{figure}

Our model can also be extended to facial wearable devices by integrating additional sensory inputs into the audio-based diffusion framework. 
We demonstrate this with a VR headset with two head-mounted cameras (HMC) near the eye regions and a microphone. 
This multimodal approach addresses the limitations of head-mounted cameras that struggle with lower face capture due to occlusions. 
This setup is also applicable to lightweight devices such as smart glasses where additional visual sensors beyond eye cameras are hardware-wise infeasible.
In this setting, the gaze vector $\mathbf{g}$ is replaced by eye features $\mathbf{e} \in \mathbb{R}^{160}$. These features were extracted from HMC eye images (monochrome images of resolution $400 \times 400$) using the feature extraction part of the facial encoding system presented in~\cite{bai2024ue}. For an overview of the system, see Fig.~\ref{fig:mm_sys}.

\section{Experiments}
\label{sec:experiments}

\begin{table*}
\small
    \centering
    \caption{\textbf{Quantitative comparison with state-of-the-art baselines.} 
    Our method shows overall superior performance even compared to offline baselines that rely on future information, while being over $1000\times$ faster than the slowest baseline.}
    \begin{tabular}{l|l|c|c|c|c|c|c|c|c}
    \toprule
    Methods & Condition & Online & FPS (GPU Time) $\uparrow$  
    & \multicolumn{3}{c|}{\textit{Dataset: freeform}} 
    & \multicolumn{3}{c}{\textit{Dataset: sentence}} \\
    & & & & LVE $\downarrow$ & FDD $\downarrow$ & Lip Sync $\downarrow$ 
            & LVE $\downarrow$ & FDD $\downarrow$ & Lip Sync $\downarrow$ \\
    \midrule
    TalkShow-Face & Audio & \xmark & 
    133 (7.5 ms) & 
    \cellcolor{orange!20}6.423 & \cellcolor{orange!20}0.255 & \cellcolor{orange!20}5.114 &
    \cellcolor{orange!20}5.541 & \cellcolor{orange!35}0.132 & \cellcolor{orange!20}4.583 \\
    Audio2Photoreal-Face & Audio & \xmark & 
    0.77 (1.3 sec) & 
    8.490 & 0.259 & 6.109 & 
    6.298 & 0.167 & 4.897 \\
    DiffPoseTalk w/o Style & Audio & \xmark & 
    0.22 (4.5 sec) & 
    11.506 & 0.579 & 9.515 & 
    10.805 & 0.506 & 9.254 \\
    \textcolor{gray}{DiffPoseTalk w/ Style}  & \textcolor{gray}{Audio + Style (from GT exp.)} & \textcolor{gray}{\xmark} & 
    \textcolor{gray}{0.09 (11.4 sec)} & 
    \cellcolor{orange!35} \textcolor{gray}{6.421} & 
    \cellcolor{orange!70} \textcolor{gray}{0.161} & 
    \cellcolor{orange!35} \textcolor{gray}{4.774} &
    \cellcolor{orange!35} \textcolor{gray}{5.596} & 
    \cellcolor{orange!70} \textcolor{gray}{0.102} & 
    \cellcolor{orange!35} \textcolor{gray}{4.322} \\
    \textbf{Ours} & Audio + Gaze (synthesized) & \checkmark & 
    \textbf{100 (10 ms)} & 
    \cellcolor{orange!70} \textbf{6.329} &    
    \cellcolor{orange!35} \textbf{0.185} & 
    \cellcolor{orange!70} \textbf{4.751} &
    \cellcolor{orange!70} \textbf{5.177} & 
    \cellcolor{orange!20} \textbf{0.146} & 
    \cellcolor{orange!70} \textbf{4.178} \\
    \bottomrule
    \end{tabular}
    \label{tab:baseline_quant}
\end{table*}

\subsection{Experimental Settings}
\label{sec:exp_settings}

\paragraph{Data.} 
Our in-house dataset consists of two sets; the first set 
is a collection of 265 subjects whose facial expressions and audio were captured in a multi-view capture system, following~\cite{li2024uravatar}. 
During the capture, each subject speaks 5 minutes of \textit{freeform} speech, and reads 30 to 35 \textit{sentences} where each sentence lasts about 3 seconds. 
The audio and corresponding facial expressions are used to train the audio and gaze-conditioned diffusion model $\mathcal{E}$, including distillation.
\camr{
The second set used for emotion conditioning application (Sec.~\ref{sec:emo_cond}) has the same number of subjects as the dataset used for audio-driven encoder model training, where
each subjects are captured with 8 head mounted cameras (HMC) using an augmented VR headset. 
The ground truth expression codes are extracted from the HMC captured images using the framework of ~\cite{bai2024ue, wei2019VR}. 
Each subject is asked to read sentences and give a short freeform speech with emotional expressions.
The dataset have 10 emotion labels in total, and per each label, sentence reading and short speech last about 11 seconds and 15 seconds respectively. 
}

\indent\textit{Metrics.} We compare our method with baselines with the following well-established metrics.
(1) \textbf{Lip Vertex Error (LVE)} (mm): Measures lip movement deviation by computing the maximum $L_2$ error across all lip vertices per frame, averaged over all frames.
(2) \textbf{Upper-Face Dynamic Deviation (FDD)} (mm): Evaluates upper face motion by comparing standard deviations of upper face vertex movements.
(3) \textbf{Lip Sync} (mm): Measures lip synchronization accuracy by calculating horizontal and vertical $L_2$ distances between lip corner pairs and upper-lower lip centers, then summing up the deviations.
(4) The naturalness of the generated expressions are evaluated through a perceptual study (A/B test). 
We measure the selection ratio compared to ours.

\indent\textit{Latency Measurement Setup.}
Quantitative experiments and latency measurements for our model and all baselines are done on \textbf{identical} GPU hardware setup, with a single NVIDIA A100 GPU. 
In the live real-time demo in the supplementary video, we used Meta Quest Pro with PyTorch’s native JIT scripting, and a consumer level NVIDIA 3090 GPU for inference. 

\begin{table}
\small
\centering
\caption{\textbf{Ablation on real-time live-driving system design} for maintaining temporal consistency and accuracy when sampling is done at every time frame $t$.
}

\begin{tabular}{l|cc|c|c|c}
\toprule
Dataset & wav2vec 1.0 & outpaint & LVE $\downarrow$ & Lip Sync $\downarrow$ & Lip Vel $\downarrow$
\\ \midrule
\textit{freeform}  & & & 7.433 & 6.045 & 1.758 \\
& \(\checkmark\) & & 
\cellcolor{orange!35} 7.100 & 
\cellcolor{orange!35} 5.556 & 
\cellcolor{orange!35} 1.252 \\
& \(\checkmark\) & \(\checkmark\) & 
\cellcolor{orange!70} \textbf{6.966} & 
\cellcolor{orange!70} \textbf{5.426} &
\cellcolor{orange!70} \textbf{1.245} \\
\midrule
\textit{sentence} & & & 6.444 & 5.342 & 1.619 \\
& \(\checkmark\) & & 
\cellcolor{orange!35} 6.048 & 
\cellcolor{orange!35} 5.010 & 
\cellcolor{orange!35} 1.102 \\
& \(\checkmark\) & \(\checkmark\) & 
\cellcolor{orange!70} \textbf{5.936} & 
\cellcolor{orange!70} \textbf{4.992} &
\cellcolor{orange!70} \textbf{1.062} \\
\bottomrule
\end{tabular}
\label{tab:abl_realtime}
\end{table}

\begin{table}
\small
    \centering
    \caption{\textbf{Ablation study results} on using distillation with diffusion for single step generation.}
    \begin{tabular}{l|l|c|c|c}
    \toprule
    Dataset & Methods & LVE$\downarrow$ & FDD$\downarrow$ & Lip Sync$\downarrow$ \\
    \midrule
    \textit{freeform} & Regression & 7.477 & 0.208 & 5.852 \\
    & Ours (Diffusion+Distill) & \textbf{6.589} & \textbf{0.179} & \textbf{4.968} \\
    \cmidrule{1-5}
    \textit{sentence} & Regression & 6.398 & 0.137 & 5.034 \\
    & Ours (Diffusion+Distill) & \textbf{5.021} & \textbf{0.122} & \textbf{3.969} \\
    \bottomrule
    \end{tabular}
    \label{tab:abl_single_step}
\end{table}

\subsection{Baseline Comparison}
\label{sec:baseline_comp}
\paragraph{Baselines} 
Since we present the first online real-time model to universally drive high-fidelity 3D faces from audio, there exists no direct online-based competitor with the same setup.
\camr{While other baselines such as GaussianTalker~\cite{cho2024gaussiantalker} and TalkingGaussian~\cite{li2025talkinggaussian} generate high-fidelity 3DGS face animation driven from audio, we do not set those as direct baselines as they directly learn person-specific 3D deformation and rendering from audio. In contrast, our method targets learning universal and identity-agnostic latent facial expressions through an encoder model, which are consequently transformed into 3DGS-based facial avatars based on identity conditions. Given these fundamental differences in approach and scope,}
we compare against the most relevant offline state-of-the-art baselines that generate latent facial expressions that are decoded into 3D to compare the performance on learning the sequential distribution of facial expressions.  

\begin{itemize}
    \item \textbf{DiffPoseTalk}~\cite{sun2024diffposetalk}
is a state-of-the-art method which outperforms key prior works such as FaceFormer~\cite{fan2022faceformer} and CodeTalker~\cite{xing2023codetalker} by employing audio and style conditioned diffusion models,
where style is first extracted via a transformer-based style encoder from expressions. 
Here we compare with two settings: a \textit{hypothetical setting} where style is extracted from \textit{ground truth} expressions (which is not provided in real-world demonstrations) and a setting without style conditioning.
    \item \textbf{TalkShow}~\cite{yi2023talkshow} uses a convolution-based regression model to regress facial expressions from audio. 
    \item \textbf{Audio2Photoreal}~\cite{ng2024audio2photoreal} utilizes diffusion models to generate facial expressions from conversational audio.
It also uses a pretrained lip regressor which regresses lip geometry from audio as an additional conditional input. For our setting, we change two person audio input to a single person (speaker) audio and use the identical pretrained lip regressor model.

\end{itemize}

\begin{figure*}
    \centering
    \includegraphics[width=0.9\linewidth, trim={0 0.5cm 0 0}]{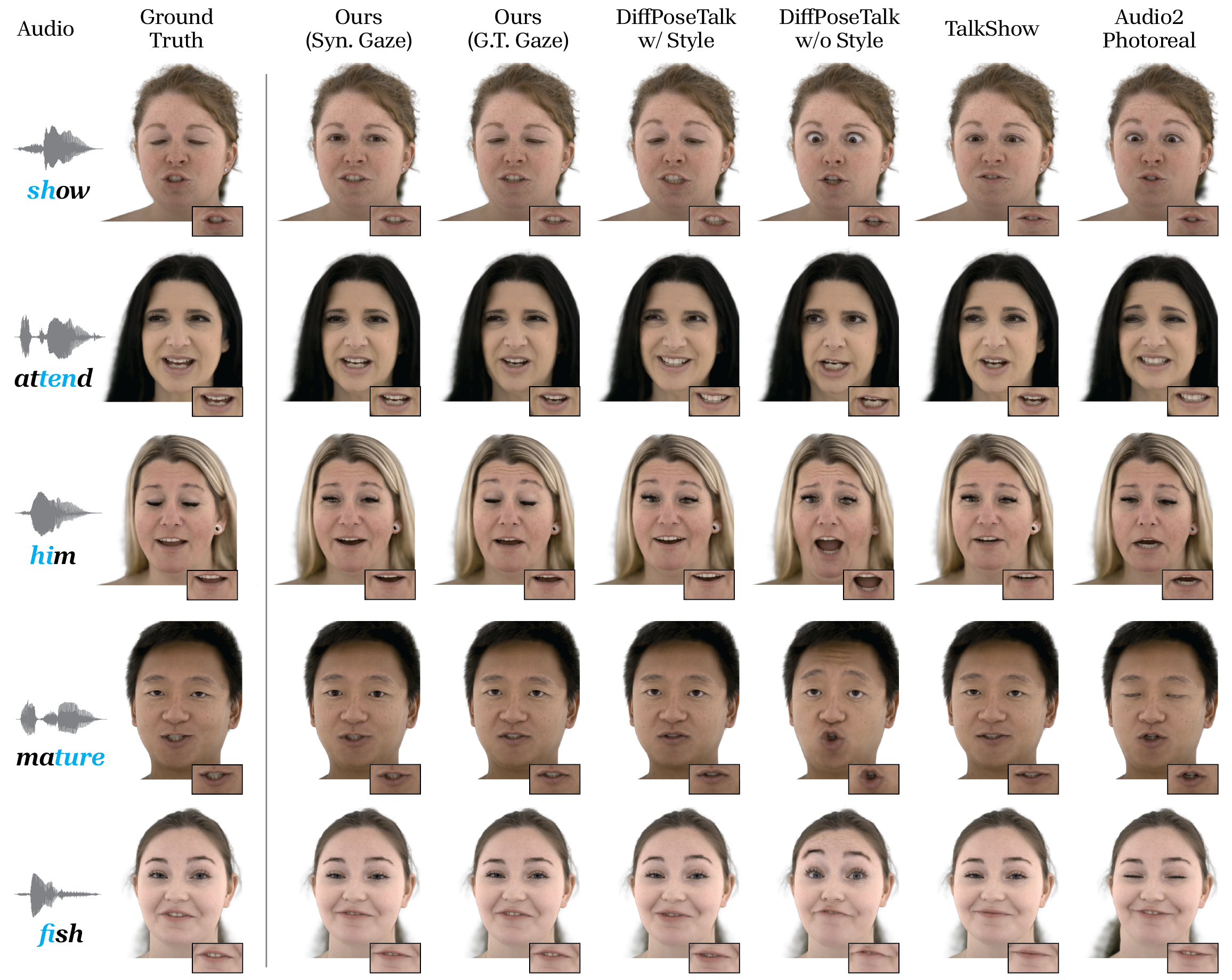}
    \caption{\textbf{Visual comparison with state-of-the-art baselines. } Compared to baselines, our method generates natural facial expressions with synchronized lip movements while meeting stricter latency constraints.}
    \label{fig:baseline_comp}
\end{figure*}
\begin{figure*}
    \centering
    \includegraphics[width=0.9\linewidth, trim={0 0.5cm 0 0}]{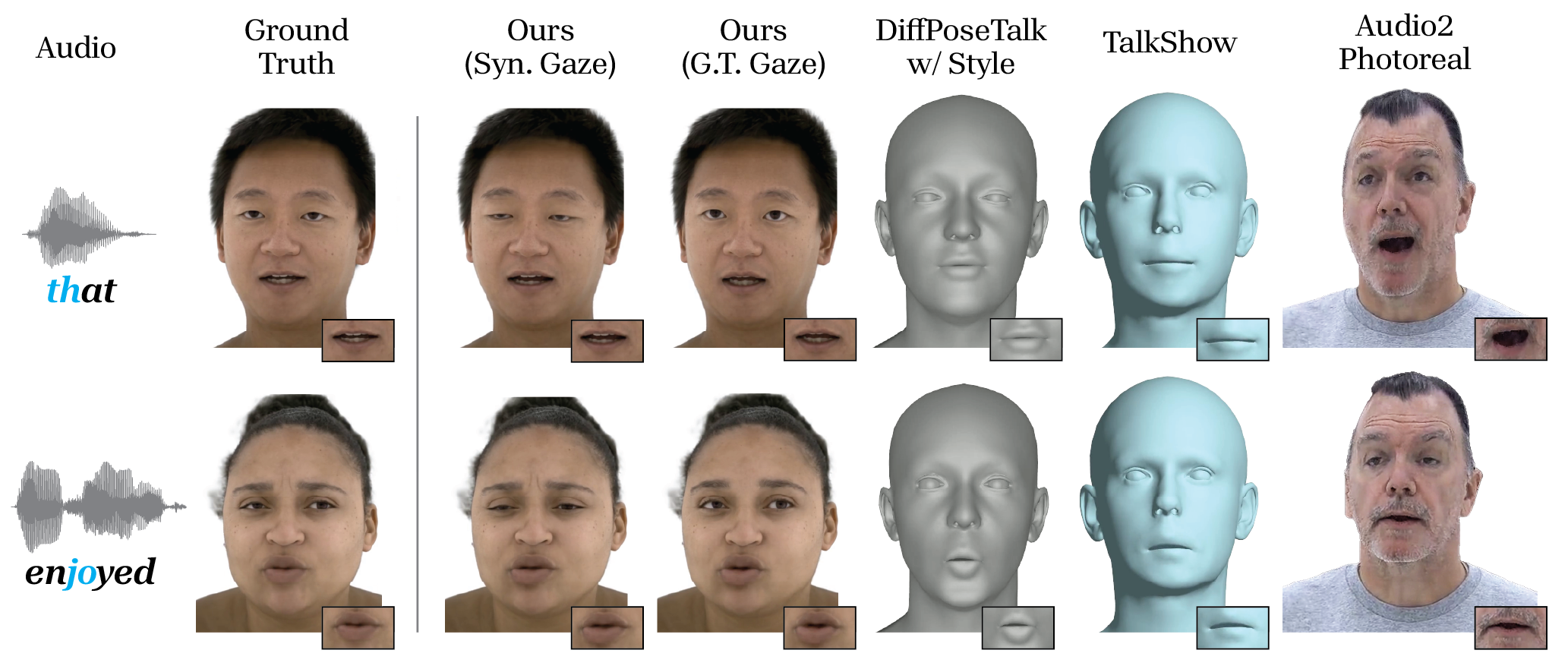}
    \caption{\textbf{Additional visual comparison with the state-of-the-art baselines. } 
    Our model can generate more accurate and detailed facial motions (e.g., tongue movements and eyebrows) compared to baselines.
    }
    \label{fig:baseline_comp_originalvert}
\end{figure*}

\paragraph{Results.}
For fairness, we follow the official implementation of baselines and the expressions of all methods are decoded with an identical decoder model.
Also, we follow the settings of DiffPoseTalk~\cite{sun2024diffposetalk} for the sampling procedure (5 samples) and sequence length for evaluation.

As shown in Table~\ref{tab:baseline_quant}, our method achieves \textbf{real-time performance} ($<$ 15ms) and operates \textbf{online}, yet outperforms offline baselines that rely on future information.
Also, as visually demonstrated in Fig.~\ref{fig:baseline_comp}, our method generates natural and synchronized lip and facial movements compared to baselines.
Even under a hypothetical setting for DiffPoseTalk (w/ Style), where style conditions are derived from ground truth expressions, our method achieves superior or comparable results with \textit{over a $1000\times$ speed improvement}.
The relatively high FDD is due to our model being conditioned on both audio and gaze. When the synthesized gaze differs from ground truth, our model generates facial expressions aligned with the given synthesized gaze input, as seen in the second and third column of Fig.~\ref{fig:baseline_comp}. 
This leads to discrepancies in upper-face behavior compared to ground truth, particularly in sentence datasets where ground truth upper-face movements are minimal.
With ground truth gaze, our model achieves FDD scores of 0.179 for freeform and 0.122 for sentence reading, outperforming TalkShow and comparable to the hypothetical DiffPoseTalk (w/ Style) baseline.

\camr{For a more comprehensive comparison, we also compare with the original version of the baseline methods that output 3DMM and FLAME based facial meshes (DiffPoseTalk, TalkShow) and personalized photorealistic avatars (Audio2Photoreal), where the results are showin in Fig.~\ref{fig:baseline_comp_originalvert}. 
For the dental fricative sound `th' in `that' (Fig.~\ref{fig:baseline_comp_originalvert} top), our model generates detailed lip movements with the tongue positioned between the teeth to produce the fricative sound. While DiffPoseTalk produces similar overall lip shapes, it fails to capture the precise teeth and tongue details characteristic of fricative sounds. TalkShow and Audio2Photoreal exhibit inaccurate lip synchronization and motion compared to the ground truth. Similarly, when pronouncing `joy' in `enjoyed' (Fig.~\ref{fig:baseline_comp_originalvert}, bottom) our model reproduces fine lip details, whereas DiffPoseTalk achieves comparable lip synchronization but lacks detailed articulation. TalkShow and Audio2Photoreal again demonstrate inaccurate lip motion.
}

\paragraph{Perceptual Study.}
We also perform a perceptual study with results of our method and baselines where all the results are decoded with an identical decoder model.
Compared to TalkShow, 75.8\% of users favored our method; to DiffPoseTalk, 84.59\%; and to Audio2Photoreal, 69.17\%. 
These results demonstrate that our method generates more natural facial expressions, even under the constraint of online and real-time inference. 

\begin{figure}
    \centering
    \includegraphics[width=0.85\linewidth, trim={0 0.5cm 0 0}]{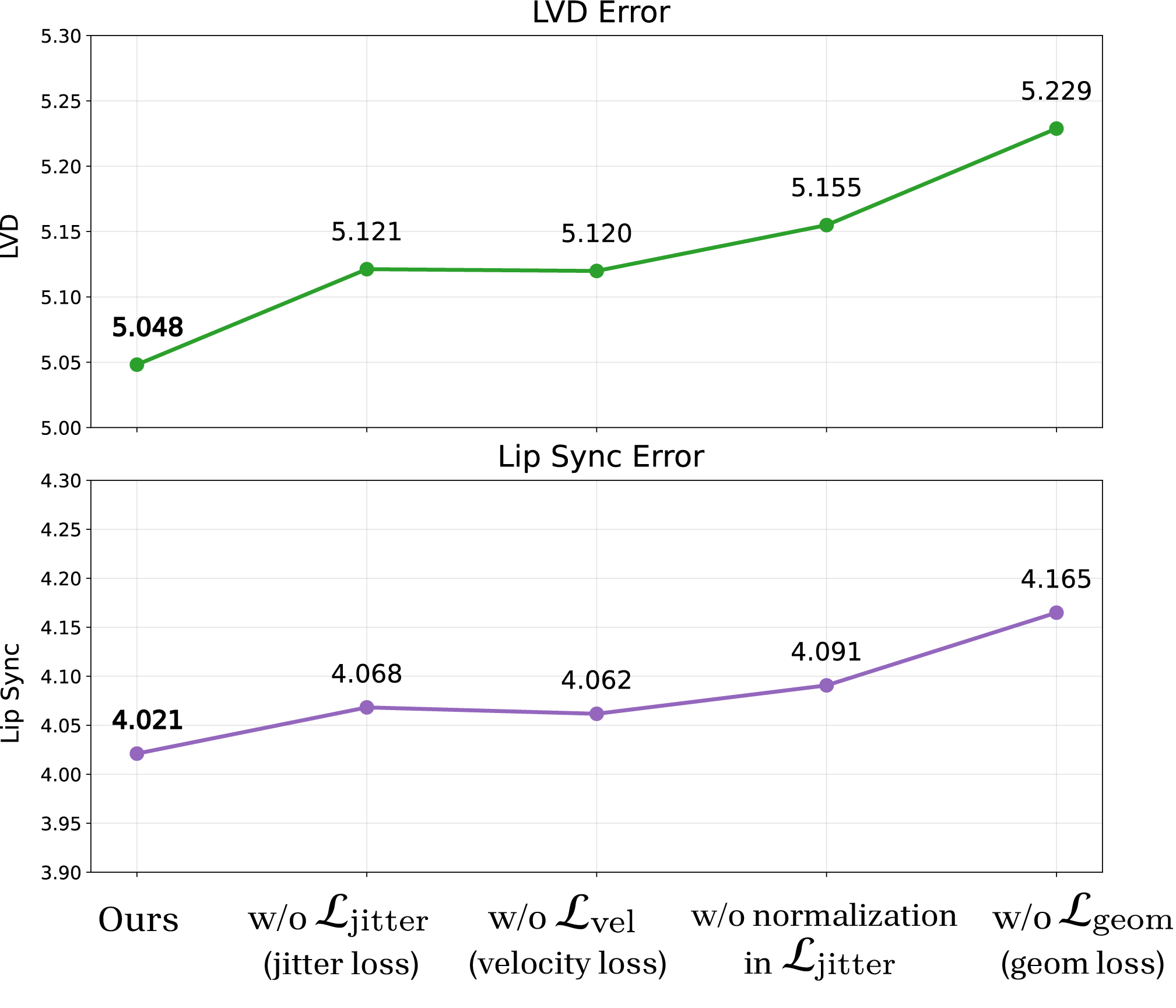}
    \caption{\textbf{Ablation studies on each component $\mathcal{L}_\text{vel}$ and $\mathcal{L}_\text{vel}$  of geometry loss $\mathcal{L}_{\text{geom}}$.}
    Both velocity loss and jitter loss components contribute significantly to learning accurate facial (lip) movements, with normalization in jitter loss playing a critical role.
    }
    \label{fig:geom_compare}
\end{figure}

\begin{figure}
    \centering
    \includegraphics[width=0.85\linewidth, trim={0 0.4cm 0 0.2cm}]{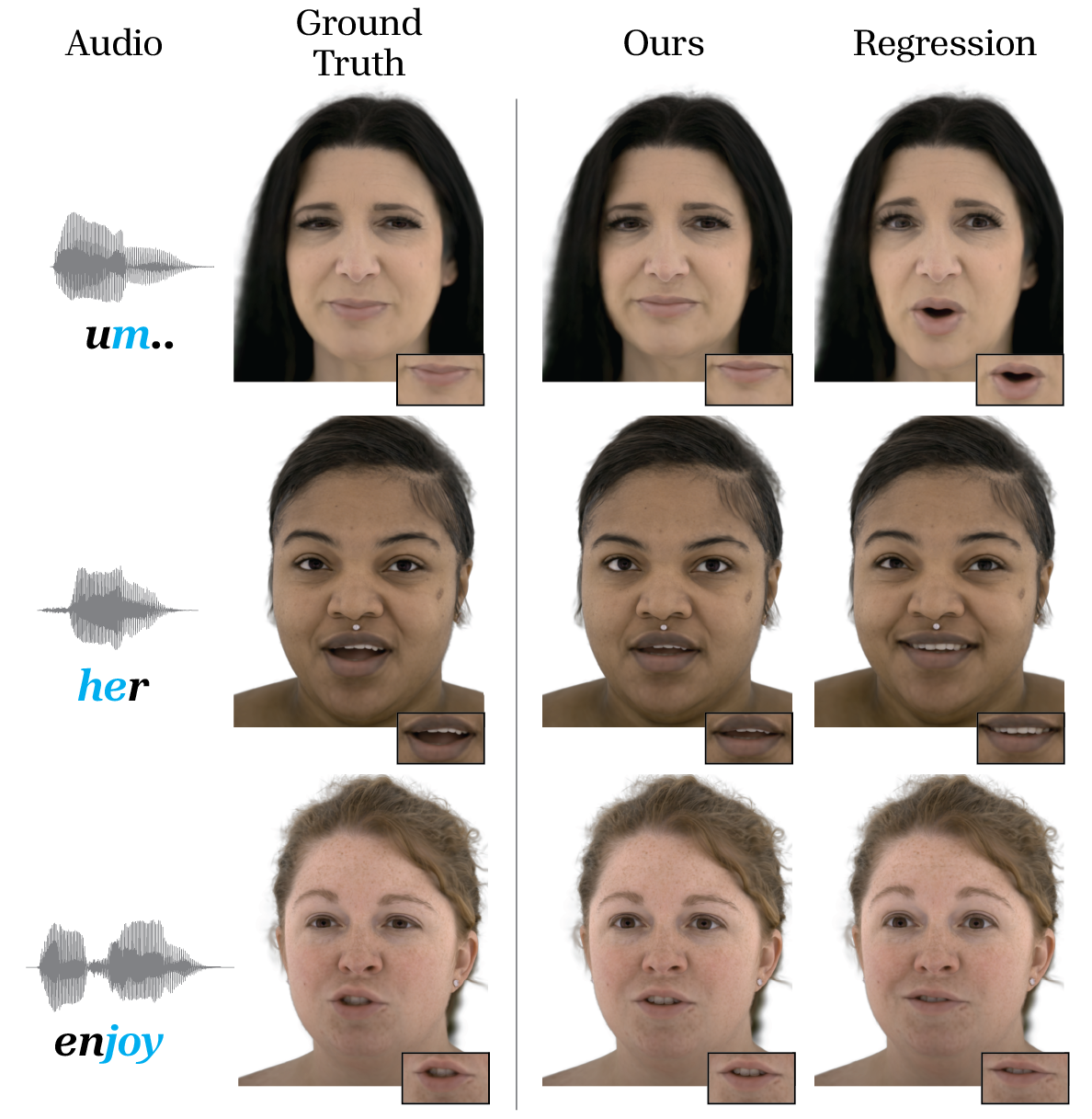}
    \caption{\textbf{Visual comparison of our method  with regression baseline for single step generation.} 
    For single step inference, combining distillation with diffusion generates more natural and accurate lip motions
    compared to the regression baseline.
    }
    \label{fig:reg_comp}
\end{figure}

\begin{figure}
    \centering
    \includegraphics[width=0.85\linewidth, trim={0 0.2cm 0 0.3cm}]{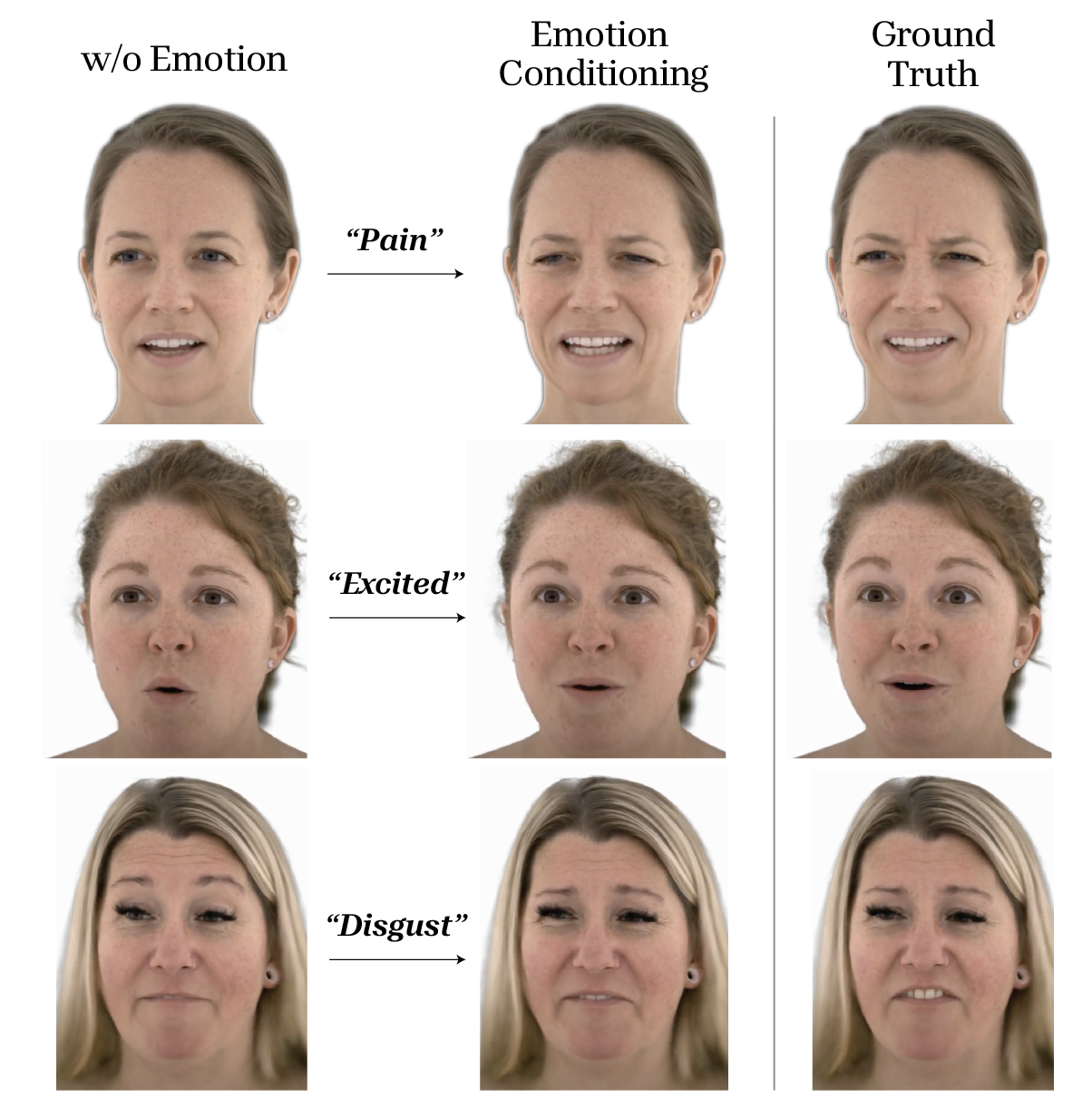}
    \caption{\textbf{Visual comparison of facial expressions with and without emotion conditioning.} Emotion conditioning model $\mathcal{E}_\text{emo}$ generates expressions that matches the given emotional context.
    such as frowning (``pain", ``disgust") or widening eyes and raising eyebrows (``excited").
    }
    \label{fig:emo_cond}
\end{figure}

\begin{figure}
    \centering
    \includegraphics[width=1.0\linewidth, trim={0 0.3cm 0 0}]{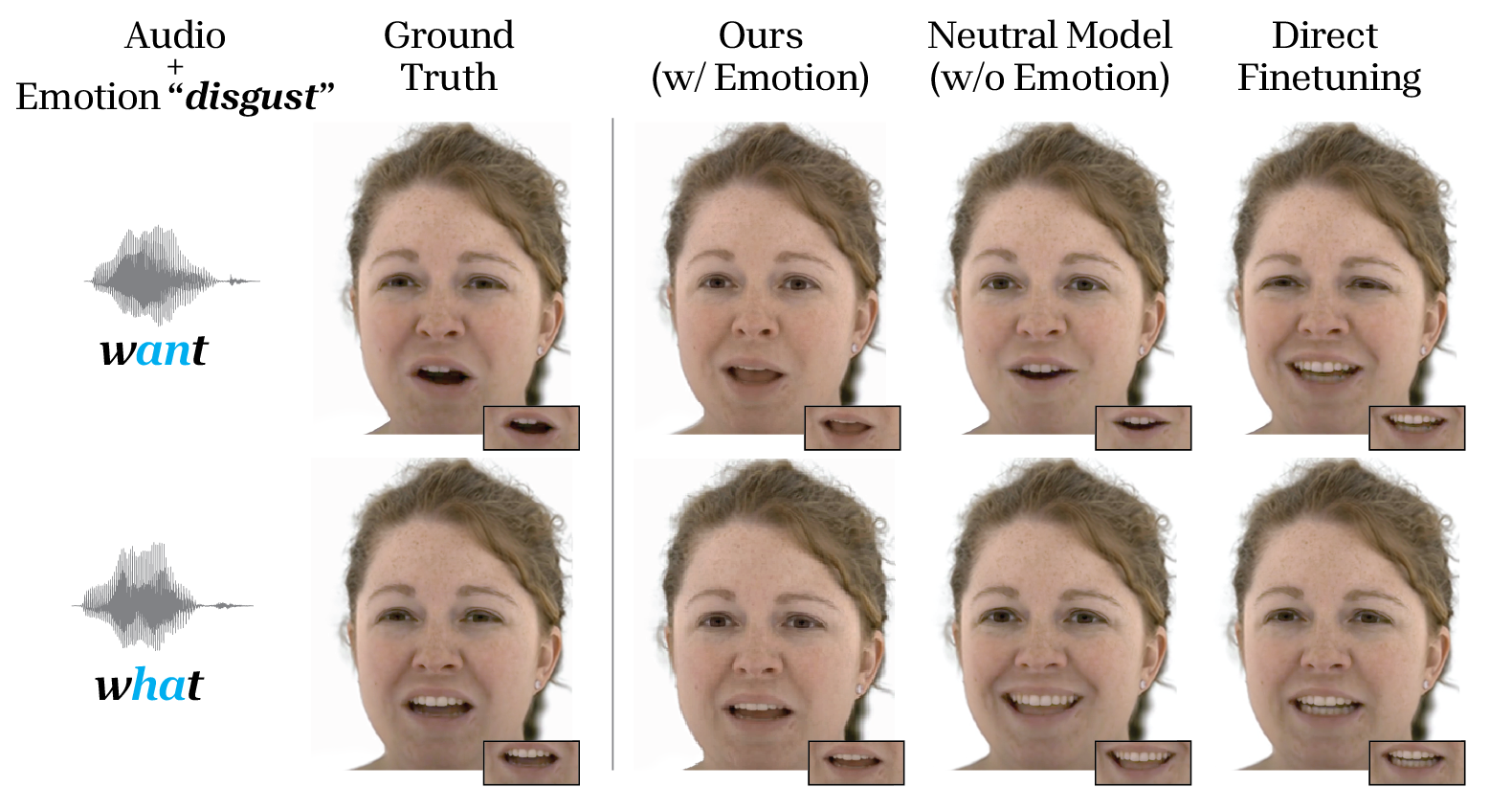}
    \caption{
\textbf{Ablation of emotion conditioning design.}
Adding zero-initialized conditional layers improves alignment with emotion labels over the neutral model and preserves more accurate lip movements compared to direct finetuning.
}
\label{fig:emo_abl}
\end{figure}

\subsection{Ablations}
\label{sec:abl}
We perform ablation studies to assess the contributions of each component of our method. 

\paragraph{Analysis on Geometry Loss.}
We quantitively analyze the contribution of geometry loss $\mathcal{L}_\text{geom}$ for the audio and gaze conditioned diffusion model to learn natural facial motions in online settings (Sec.~\ref{sec:audio_diffusion}).
The graph in Fig.~\ref{fig:geom_compare} shows the individual contributions of the velocity and jitter loss components, as well as the effect of jitter normalization, in learning an accurate temporal facial dynamics.

\paragraph{Single Step Generation.}
For single step inference (Sec.~\ref{sec:singlestep}), we compare ours to regression with the same architecture and input. 
Table~\ref{tab:abl_single_step} and Fig.~\ref{fig:reg_comp} shows distillation outperforms regression,
highlighting the strength using the prior learned through diffusion. 

\paragraph{Live-Driving System Design.}
We evaluate the efficacy of our live driving system components by simulating a live driving scenario
where noise is differently sampled for every timestep $t$ and facial expressions are generated frame-by-frame.
We compare the causal audio encoder (wav2vec 1.0~\cite{schneider2019wav2vec1}) against non-causal encoder (wav2vec 2.0~\cite{baevski2020wav2vec2}) and assess the impact of our simplified outpainting approach. Table~\ref{tab:abl_realtime} shows that combining both causal encoding and outpainting produces temporally consistent motions with lower lip velocity error while maintaining accuracy.
These results validate our design choices for real-time applications, addressing constraints not typically considered in previous state-of-the-arts.

\paragraph{Design for Emotion Conditioning Application.} 
To assess the impact of emotion conditioning (Sec.~\ref{sec:emo_cond}), we compare expressions generated by the neutral model $\mathcal{E}$ and the emotion-conditioned model $\mathcal{E}_\text{emo}$ in Fig.~\ref{fig:emo_cond}. Fig.~\ref{fig:emo_abl}
further shows the effectiveness of utilizing zero-initialized conditional layers compared to direct finetuning for accurate lip motions with emotional expressions.

\section{Discussion}
Our system achieves social telepresence by driving high-fidelity 3D facial avatars with audio in real-time. Through our online transformer architecture and distillation pipeline, we accelerate diffusion model inference to a single step, enabling real-time performance. We also address system design challenges for robust live-driving, and demonstrate multimodal extensions such as emotion labels and additional sensors in VR headsets.

\camr{
\paragraph{Limitations.}
Our approach has several limitations. First, in real-time live-driving scenarios where audio signal injection and inference is done every frame, while outpainting preserves consistency, some jitter may remain.
Second, although the decoder model achieves high-fidelity rendering of facial regions where expressions are most prominent, several artifacts may exist in hair and oral cavity interiors such as teeth and tongue. 
}

\paragraph{Future Work.}
Our method achieves real-time performance with a single GPU, and we leave model quantization for on-device computation as a future direction.
Due to the lack of head pose information in the current dataset, our pipeline does not model it explicitly. Extending the pipeline to include head pose would be an interesting future direction. 

\paragraph{Societal Impact.}
Regarding societal impact, audio-driven real-time facial avatar animation technology offers significant benefits for accessibility and immersive applications but poses risks. Responsible development requires implementing robust safeguards to maximize societal value while minimizing potential harm.

\begin{acks}
The work of Lee and Joo is supported in part by NRF grant funded by the Korean government (MSIT) [No. RS-2022-NR070498], and IITP grant funded by the Korea government (MSIT) [No. RS-2024-00439854]. 
\end{acks}

\bibliographystyle{ACM-Reference-Format}
\bibliography{main}

\clearpage
\appendix
\renewcommand{\thesection}{\arabic{section}} 
\numberwithin{equation}{section}
\setcounter{equation}{0}
\numberwithin{figure}{section}
\setcounter{figure}{0}
\numberwithin{table}{section}
\setcounter{table}{0}


\section{Video Qualitative Results}
\subsection{Live Driving Demo} In the supplementary video we showcase a live demo system that integrates all our contributions, enabling high-fidelity, real-time control of photorealistic avatars driven by unseen subjects' audio input. We also demonstrate live driving demonstrations in multimodal settings as described in Sec. 5.2, where eye HMC images and audio are given as input.

\subsection{Diverse Audio Inputs} In the supplementary video, we include visualization of the facial expressions generated from various audio inputs. We first include the results tested on the \textit{freeform speech} dataset used in the quantitative experiments. To further evaluate the robustness and generalization ability of our model, we further test our model on \textit{out-of-distribution} audio, such as speaking in multiple languages. \textbf{All the results are generated by simulating real-time live driving, where audio input is fed frame by frame and our model produces corresponding facial expressions in real time.}

\subsection{Multimodal Applications} For emotion modality settings, we present visual results in the supplementary video for cases with and without emotion modalities (Sec. 5.1). While adding emotion modality modifies the original facial expressions to match the emotional context, lip movements are preserved to stay synchronized with audio. We also show live driving demonstrations in multi-sensor settings with a VR headset (Sec. 5.2) in the supplementary video.

\subsection{Baseline Comparison} In addition to quantitative comparison, we also qualitatively compare the visual quality of the audio-driven facial expressions derived from our method and baselines in the supplementary video. We compare with state-of-the-art baselines  (Sec. 6.2) with ours and with regression baseline for single step generation (Sec. 6.3).

\subsection{Additional Experimental Results}
Along with comparsion with state-of-the-art baselines, in the additional video we visually demonstrate experimental results on (1) diverse facial expressions generated from a single audio input; (2) visual comparison with original model and single step distilled model; 
(3) comparison with regression baseline on single step generation; (4) comparison with 2D portrait generation methods.

\section{Additional Experiments}
\label{sec:add_exp}

\subsection{Comparison with Geometry Based Methods}
Although not direct baselines, we also compare with geometry-based prior works FaceFormer~\cite{fan2022faceformer}, CodeTalker~\cite{xing2023codetalker}, and Imitator~\cite{thambiraja2023imitator} that directly learn geometric deformations of coarse facial meshes with fixed topology. 


In our latency comparison (conducted using identical hardware configurations and measuring GPU time, as in the main paper), although these methods are online, they are shown to be unsuitable for real-time applications due to computational latency as their networks directly output geometry deformations from audio features. 
Our approach demonstrates advantages by substantially reducing computational overhead during single inference operations while generating facial animation with higher fidelity and much finer details than coarse facial mesh-based methods. 
This is achieved by representing facial motion as latent expression sequences and transforming them into photorealistic Gaussian-based faces through a decoder, resulting in up to 100$\times$ speed improvement.
Furthermore, our method ensures causality in the audio encoder, enabling fully real-time operations which is a distinct advantage over existing approaches.

\begin{table}
\small
\footnotesize
\caption{\textbf{Additional latency comparison with existing mesh-based methods.} Our method achieves the lowest latency and also ensures full causality.}
\begin{tabular}{l|c|c|c}
\toprule
\textbf{Method} & Latency / FPS & Audio Encoder Causality & Online \\
\midrule
FaceFormer & 309ms (3.2FPS) & \xmark & Autoregressive \\
CodeTalker & 1250ms (0.8FPS) & \xmark & Autoregressive \\
Imitator & 228ms (4.38FPS) & \xmark & Autoregressive \\
Ours & \textbf{10ms (100FPS) }& \checkmark & Windowed Attention \\
\bottomrule
\end{tabular}
\label{tab:latency_comp_vertical}
\end{table}

\subsection{Comparison with 2D Based Methods}

To demonstrate the advantages of our 3D-based approach, we compare it with 2D-based methods that generate portrait image sequences synchronized with audio. We use AniPortrait~\cite{wei2024aniportrait} as a baseline, which leverages 3D information extracted from images as an intermediate representation and image diffusion models to synthesize natural portrait images. 
As shown in Fig.~\ref{fig:2d_comp} and in the supplementary video, driving 3D facial avatars offers significant advantages in generating temporally consistent and 3D-plausible results. AniPortrait's results exhibit blurry teeth artifacts and distorted lip shapes that are spatiotemporally inconsistent, while our method generates natural, high-fidelity avatars with accurately synchronized lip movements.
\begin{figure}
    \centering
    \includegraphics[width=0.9\linewidth, trim={0 0.2cm 0 0}]{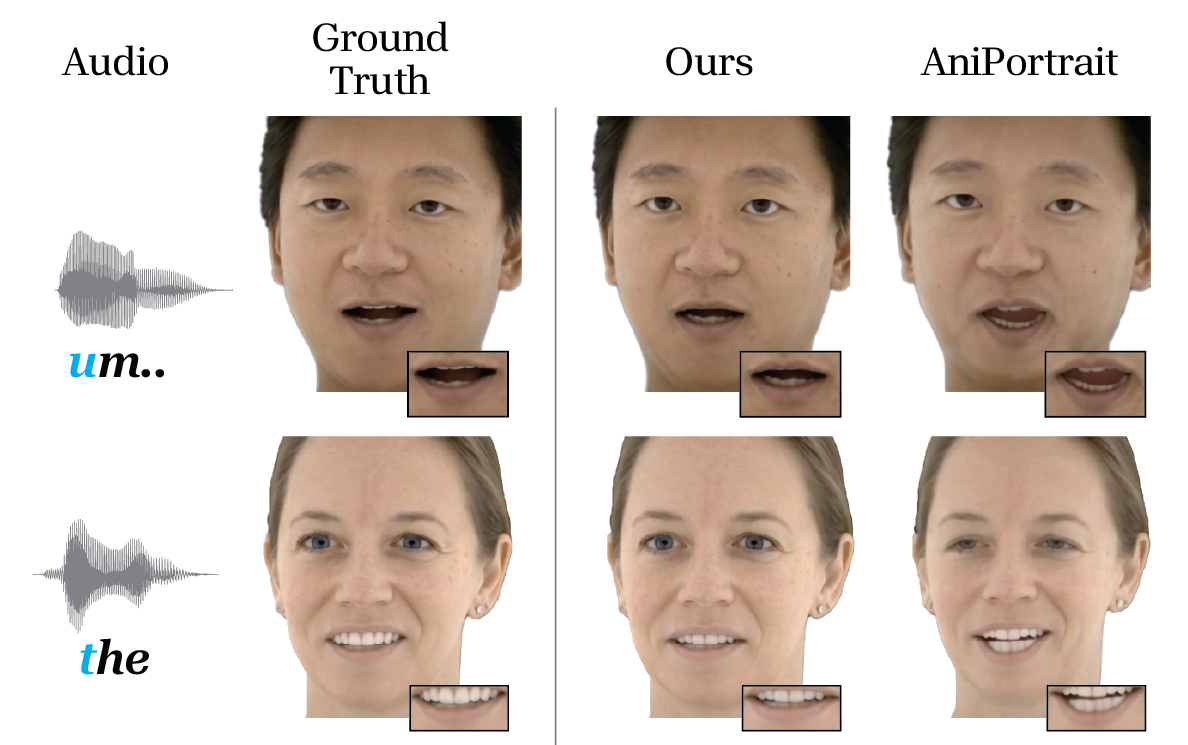}
    \caption{
\textbf{    Visual comparison of our method and 2D-based methods.}
 We compare with AniPortrait~\cite{wei2024aniportrait} as a baseline.
AniPortrait's results show blurry and distorted artifacts, while our method generates natural and detailed avatars with accurate lip movement. Refer to Supp. Sec.~\ref{sec:add_exp} for more details.}
    \label{fig:2d_comp}
\end{figure}

\section{Implementation Details}

\subsection{Online Transformer Architecture}
The self-attention operation $\mathcal{A}(\mathbf{Q}, \mathbf{K}, \mathbf{V})_t$ for the $t$-th frame with RoPE and windowed mask can be defined as:
\begin{equation}
    \mathcal{A}(\mathbf{Q}, \mathbf{K}, \mathbf{V})_t =
    \frac{
        \sum_{i=t-w}^{t} \mathbf{R}_t \phi(\mathbf{q}_t)^\top \mathbf{R}_i \varphi(\mathbf{k}_i) \mathbf{v}_i
    }{
        \sum_{i=t-w}^{t} \phi(\mathbf{q}_t)^\top \varphi(\mathbf{k}_i)
    }
    \label{eq:attn}
\end{equation}
where $\mathbf{q}_i, \mathbf{k}_i, \mathbf{v}_i \in \mathbb{R}^D$ represent the query, key, and value vectors for the $i$-th input token while $\mathbf{Q}, \mathbf{K}, \mathbf{V}$ denote the concatenated query, key, and value matrices.
The feature projection functions $\phi(\cdot)$ and $\varphi(\cdot)$ are applied to the query and key, followed by a multiplication with the rotary matrix $\mathbf{R}_k$ which is defined based on the timestep $k$.  For details refer to the original RoPE paper~\cite{su2024roformer}. Unlike the attention function in the original paper, note that in Eq.~\ref{eq:attn}  the attention function $\mathcal{A}(\mathbf{Q}, \mathbf{K}, \mathbf{V})_t$ at time $t$ considers only the input tokens within the range $t-w$ to $t$, as enforced by the windowed mask.

Compared to conventional sliding window based approaches with fixed lengths and positional embeddings, fixing the receptive field with a windowed mask resolves boundary issues (e.g., edge token inconsistencies). Additionally, while such conventional approaches reset absolute positions for each window and thus disrupt coherence, the use of RoPE directly encodes relative positions and better maintains temporal coherence among frames.

\subsection{Gaze Synthesis}
The node $\gamma$ of the gaze graph $\mathcal{G}$ consists of gaze vector sequences $\mathbf{G}_{:w_g} = \{\mathbf{g}_t\}_{t}^{t+w_g}$ . 
The distance of directional edge from node $\gamma_i$ to $\gamma_j$ are based on the position and velocity difference between the last $w_f$ frames in node $\gamma_i$ and the first $w_f$ frames in node $\gamma_j$.
Formally, the distance can be calculated as:
\begin{equation}
    \sum_{t=0}^{w_f} \|{\mathbf{g}}_{(w_g - w_f) + t}^i - {\mathbf{g}}_t^{j}\|_2 +  \sum_{t=1}^{w_f} \|\dot{\mathbf{g}}_{(w_g - w_f) + t}^i - \dot{\mathbf{g}}_t^{j}\|_2
\label{eq:gaze_node_diff}
\end{equation}
where $\dot{\mathbf{g}}_t$ is simply computed as $\mathbf{g}_t - \mathbf{g}_{t-1}$. Superscript $i$ and $j$ denotes the nodes $\gamma_i$ and $\gamma_j$.

To ensure smooth transition between nodes, the edges are constructed if the distance is under a certain threshold.  
In the current implementation, the window length for $w_g$ and $w_f$ are set to 45 and 10, respectively. For measuring the distance, the differences are normalized by dividing by the standard deviation of gaze vector $\mathbf{g}$ across the whole graph $\mathcal{G}$. 
The gaze sequence $\mathbf{G}$ are synthesized by traversing the graph $\mathcal{G}$. Search and transition between nodes $\gamma$ takes around 1 to 2 milliseconds. 

The gaze graph $\mathcal{G}$ is constructed from the gaze sequences from the \textit{freefrom speech} data in the first dataset (Sec. 6.1). The graph implementation is done using the NetworkX library~\cite{hagberg2008networkx}, and after the graph is constructed, the size of the graph is reduced by retaining only the strongly connected components. The reduced graph consists of 291 nodes with 10191 edges. 

\subsection{Distillation Timestep Sampling} 
For stable distillation training, we adopt the timesteps sampling method presented in~\cite{chadebec2024flash}. 
Rather than training the distilled model with single step sampling from the start, we initially prioritize 4-step sampling and progressively transition to single step for stable training. To achieve this, the probability of timesteps are decided based on a probability mass function of timesteps $\pi(\tau)$ which is modeled as a mixture of Gaussians:
\begin{equation}
    \pi(\tau) = \frac{1}{\sqrt{2\pi\sigma^2}} \sum_{i=1}^K \beta_i \exp\left(-\frac{(\tau - \mu_i)^2}{2\sigma^2}\right)
\end{equation}
where $K$ is the number of uniformly spaced timesteps in the normalized timestep space $[0, 1]$, $\mu_i = i/K$ represents the mean of each Gaussian, and the variance is $\sigma = \sqrt{0.5}/K^2$. During the warm-up stage, the distribution prioritizes 4-step sampling by assigning higher weights ($\beta_i > 0$) to specific normalized timesteps [0.25, 0.5, 0.75, 1] while setting $\beta_i = 0$ for others. As training progresses, the distribution is gradually shifted to favor single step sampling by updating $\beta_i$. For the details, refer to~\cite{chadebec2024flash}. 
In our implementation, we set $K=40$, consistent with the number of DDIM denoising steps of the original diffusion model $\mathcal{E}_\text{orig}$. The $\beta_i$ values are updated every 2K iterations, and are updated 4 times until the training reaches 8K iterations. 





\begin{table}
\centering
\caption{Hyperparameter details of each module discussed in the main paper.}
\setlength{\tabcolsep}{3.5pt}
\resizebox{0.48\textwidth}{!}{\begin{tabular}{c|c|c}
\toprule
\multirow{5}{*}{Dataset}
& Audio frequency & 48kHz \\
& Audio channel & Single channel \\
& Audio augmentations & Noise, pitch \\
& Audio augmentation probability & 30\% \\
& Expression FPS & 30 \\
\cmidrule{1-3}
\multirow{14}{*}{Diffusion (Sec. 4.1)}
& Learning rate & 1e-04 \\
& Learning rate schedule & Cosine annealing \\
& Number of training steps & 250K \\
& Batch size & 256  \\
& Parameter optimizer & AdamW \\
& Diffusion timestep $N$ & 200 \\
& Diffusion noise schedule & Cosine \\
& Number of DDIM denoising step & 40 \\
& Classifier-free guidance dropout ratio & 0.2 \\ 
& Classifier-free guidance sampling weight & 1.0 \\
& window size $w$ & 25 \\
& Number of heads in transformer & 4 \\
& Dim of projected input & 512 \\
& Dim of latent embeddings & 1024 \\
\cmidrule{1-3}
\multirow{5}{*}{Distillation (Sec. 4.2)}
& Learning rate & 1e-05 \\
& Learning rate schedule & Cosine annealing \\
& Number of training steps & 25K \\
& Batch size & 16  \\
& Parameter optimizer & AdamW \\
\cmidrule{1-3}
\multirow{5}{*}{Emotion Conditioning (Sec. 5.1)}
& Learning rate & 1e-05 \\
& Learning rate schedule & Cosine annealing \\
& Number of training steps & 30K \\
& Batch size & 128  \\
& Parameter optimizer & AdamW \\
\bottomrule
\end{tabular}}
\label{tab:appendix-hyperparameters}
\end{table}

\subsection{Hyperparameters} In Supp. Table~\ref{tab:appendix-hyperparameters}, we provide hyperparameters used in training the models presented in the main paper.

\section{Discussions on Audio Encoders}
\begin{figure}
\includegraphics[width=0.83\linewidth, trim={0, 0.5cm, 0, 0}]{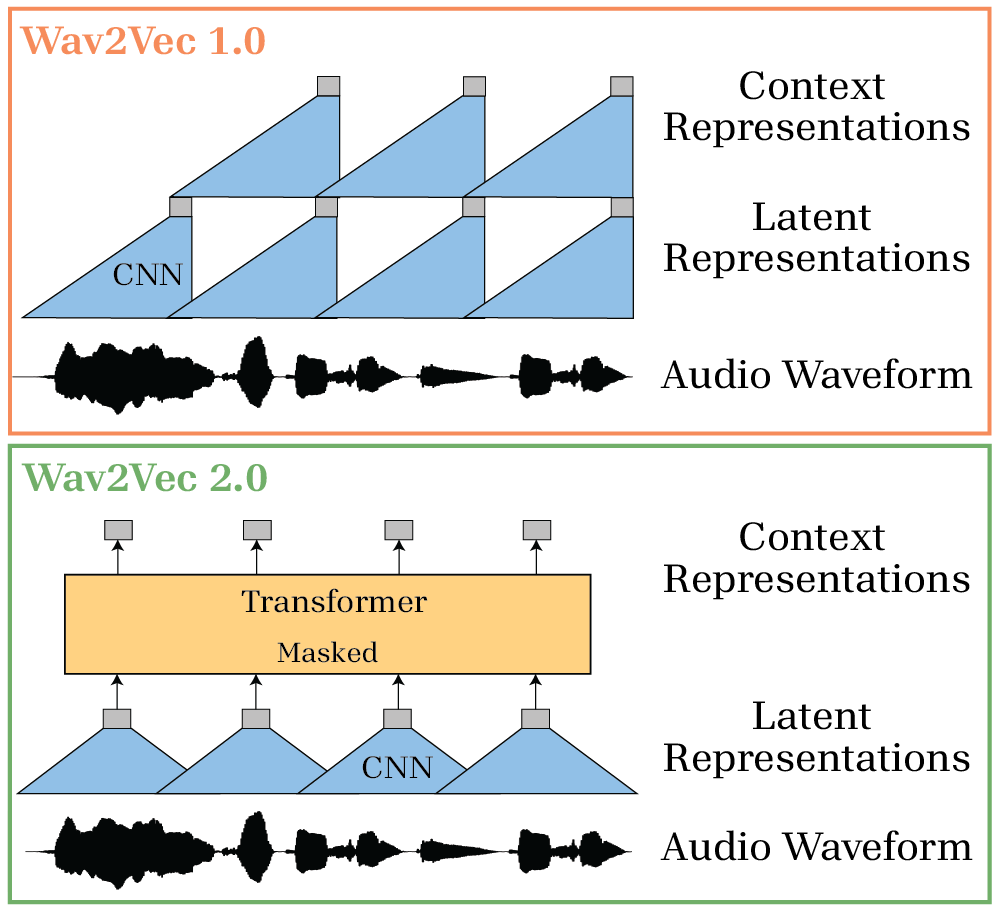}
  \caption{Visualization of network architecture of Wav2vec 1.0 and 2.0 audio encoders, adapted from the original papers.}
  \label{fig:audio_encoder}
\end{figure}

Wav2vec 1.0~\cite{schneider2019wav2vec1} employs causal convolutional neural networks (CNNs) to process raw audio waveforms into latent representations, which are converted into context representations by another layer of causal CNNs. Both CNN layers have zero lookaheads, and therefore casuality is preserved. 
In contrast, wav2vec 2.0~\cite{baevski2020wav2vec2} utilizes non-causal CNN layers, with a lookahead up to 0.5 seconds, to extract latent representations from raw audio waveforms. The latent representations undergo random masking and are consequently converted to context representations by a Transformer module. Similarly, HuBERT~\cite{hsu2021hubert} adopts the network architecture of Wav2vec 2.0, including the use of non-causal convolutions and Transformers. HuBERT additionally includes an
offline and unsupervised clustering step to generate aligned target labels. Like Wav2vec 2.0, HuBERT does not ensure full causality due to its reliance on non-causal CNN layers. 
Fig.~\ref{fig:audio_encoder} visually illustrates the network architectures of audio encoders.

\section{Experimental Settings}

%
\subsection{Data Capture and Processing}
The data in the main paper used for training audio-and-gaze-conditioned diffusion model and for distillation training are captured in a multi-view capture system equipped with 110 cameras. 
For details of the capture system setup refer to URAvatar~\citep{li2024uravatar}. 
The data used for emotion conditioning are captured by subjects wearing an augmented VR headset with 8 head mounted cameras. The ground truth expression codes are extracted from the HMC captured images using the framework of ~\cite{bai2024ue, wei2019VR}. 
For details of the data processing refer to Supp. Table ~\ref{tab:appendix-hyperparameters} (Dataset).

\subsection{Platform} 
The implementation and experiments are conducted using NVIDIA A100 GPUs with 80GB memory. The deep learning models are developed in PyTorch 2.3 with CUDA 12.1. The audio and gaze conditioned diffusion model is trained on all subjects using distributed data-parallel (DDP) using two A100 GPUs. Distillation training and emotion-conditioning training are done with a single A100 GPU. 

\subsection{Train \& Test Data Split}  
Out of 265 capture subjects, we use 237 for training and 28 for testing. 
Following baselines, data are segmented into sequences of frame length 100 (in 30FPS). 
We conduct quantitative comparison experiments on \textit{freeform speech} and \textit{sentence reading} data. For \textit{freeform speech} each subject has 70 to 90 segments, and for each subject 70 segments are randomly sampled. Out of 28 test subjects, two were excluded for excessive frame drops and less than 70 segments can be used. 
For \textit{sentence reading} each subject has 30 to 35 segments, and for each subject 30 segments are randomly sampled. Out of 28 test subjects, two were excluded for excessive frame drops and less than 30 segments can be used. 
We train our model and baseline models on identical train/test splits.

\subsection{Perceptual Study Details}
For the perception study, we prepared three test sets for each baselines: TalkShow, DiffPoseTalk, and Audio2Photoreal. 
Each test set was evaluated by a unique group of participants to ensure non-overlapping responses. 
Participants were asked to choose the facial expressions they found to be more natural and synchronized with the audio. 
Each set consisted of 24 motion sequences, and for every sequence 10 unique responses were collected, amounting to a total of 240 responses per set.

\end{document}